\documentclass[times,sort&compress]{fldauth}
\usepackage{cite}
\usepackage{moreverb}
\usepackage{lmodern}
\usepackage{graphicx}
\usepackage{subfigure}
\usepackage[english]{babel}
\usepackage{verbatim}
\usepackage{calc}
\usepackage{amsmath}
\usepackage{amssymb}
\usepackage[colorlinks,bookmarksopen,bookmarksnumbered,citecolor=black,urlcolor=black]{hyperref}
\usepackage[authoryear]{natbib}
\setcitestyle{authoryear,open={(},close={)}}

\begin{document}


\title{Investigation of Unsteady Flow Structures in a Turbulent Separating Flow using Hybrid RANS-LES Model}

\author{G. Kumar$^{\rm a}$, A. De$^{\rm a}$$^{,\ast}$ and H. Gopalan$^{\rm b}$}

\address{\centering$^{a}$Dept. of Aerospace Engineering, Indian Institute of Technology Kanpur\\ $^{b}$ Institute of High Performance Computing, Singapore}
\corraddr{Department of Aerospace Engineering, Indian Institute of Technology Kanpur, 208016, Kanpur, UP, India, E-mail:ashoke@iitk.ac.in}

\begin{abstract}
\paragraph{Purpose:}
Hybrid RANS-LES methods have become popular for simulation of massively separated flows at high Reynolds numbers due to their reduced computational cost and good accuracy. In the current study, a comparison has been made to examine the performance of Large Eddy Simulation (LES) and Hybrid RANS-LES model for a given grid resolution.\\
\paragraph{Design/methodology/approach:}
For better assessment and contrast of model performance, both mean and instantaneous flow fields have been investigated. For studying instantaneous flow, Proper Orthogonal Decomposition has been used.\\
\paragraph{Findings:}
Current analysis shows that hybrid RANS-LES is capable of achieving similar accuracy in prediction of both mean and instantaneous flow fields at a very coarse grid as compared to LES.\\
\paragraph{Originality/value:}
Focusing mostly on the practical applications of computation, most of the attention has been given to the prediction of one-point flow statistics and little consideration has been put to two-point statistics. Here, two-point statistics has been considered using POD to investigate unsteady turbulent flow.
\end{abstract}


\maketitle

\section{Introduction}


Navier-Stokes equation is an excellent mathematical model for prediction of fluid dynamics in continuum limit and a numerically accurate computation using Direct Numerical Simulation (DNS) can mimic experimental results to a high accuracy. Due to non-linearity of Navier-Stokes equation and multiplicity of scales present in the flow, it becomes computationally very expensive or sometimes impossible to compute using currently available technology and resources in many cases. This has led to the development of turbulence models to minimize computational cost while providing reasonably accurate results. For a long time, Large Eddy Simulation (LES) and Reynolds Averaged Navier-Stokes (RANS) have been very popular among fluid dynamics research and engineering community. However, the limitations of LES (high computational cost in near body region) and RANS (poor accuracy for many flow configurations) have led to the development of hybrid RANS-LES methods.

Hybrid RANS-LES models are designed to take advantage of the best features of RANS and LES. Hybrid RANS-LES models employ RANS modeling in the near-wall region and LES in the off-body region. As a result, these models combine the advantage of RANS methods in the near-wall region (lower computational cost) and LES in the region away from the wall (improved accuracy). For example, the use of hybrid methods reduces the cost of LES by 100 times at $Re=10^6$ \citep{gopalan_cost}. This reduction in computational cost will be higher as the Reynolds number is increased. In spite of this decrease in the computational cost, the accuracy of hybrid simulations has been found to be quite comparable to LES predictions \citep{abe2014advanced,breuer2008hrt,de2005hybridb,fadai2010seamless,fasel2006msc,girimaji2003ptm,gopalan_cost,hamba2003hrs,han2013efficient,hedges2002des,Menter_sstsas,shur1999des,tucker2004zonal,davidson2003hybrid}. This is possible as the size of the RANS region is relatively small compared to LES and it can be expected that it does not significantly affect the computational accuracy. This success has given rise to an increased usage of hybrid methods for simulations of flows at Reynolds number which were not possible before. A detailed review of the various types of hybrid methods and their applications can be found in Refs. \citep{frohlich2008hrm,spalart_review}.

Focusing mostly on the practical applications of computation, most of the attention has been given to prediction of one-point flow statistics and little consideration has been put to two-point statistics. Two-point velocity statistics contains information about vortical structures present in turbulent flows \citep{pope2000turbulent}. For applications in aeroacoustics and vortex induced vibration of bluff bodies, dynamics and evolution of the unsteady flows can be understood by considering two-point statistics to devise methods for active flow control, noise modeling and Reduced Order Modelling (ROM). Numerous experimental observations and DNS computation results indicate the important effect of so-called coherent structures in the flow dynamics \citep{cantwell1981organized}. Two-point statistics helps in understanding unsteady flow dynamics which considers greater importance in the prediction of flows which are not statistically stationary. This motivates us to study unsteady structures along with statistics for the development of a more economical and accurate model.  

In homogeneous fields, information can be extracted by subjecting two-point correlations to the Fourier analysis \citep{tennekes1972first}. However, most of the practically important flow cases like mixing layers, wakes or wall-bounded flows are strongly inhomogeneous. To investigate an inhomogeneous turbulent field, we have to use other approaches. 

Many turbulent flows are characterized by recurrent structures that are collectively called coherent structures. These are energetically dominant in many flows. Proper Orthogonal Decomposition (POD) is a statistical analysis method, based on the two-point correlation functions and helps in detecting coherent structures. This is a statistical technique that can be applied for the extraction of coherent structures of a turbulent flow field in fluid dynamics. The method has been independently suggested by Kosambi \citep{kosambi1943statistics}, Loeve \citep{loeve1945functions}, Karhunen \citep{karhunen1946spektraltheorie}, Pougachev \citep{pugachev1953general} and Obukhov \citep{obukhov1954statistical} and has been first introduced in turbulent flow analysis by Lumley \citep{lumley1970}. The method is extensively presented in Sirovich \citep{sirovich1987turbulence} and Berkooz et. al. \citep{berkooz1993proper}. This is a subject of great interest, as it also leads, via the Galerkin projection \citep{aubry1991hidden,panton1997self}, to a low-dimensional set of ordinary differential equations governing the evolution of vortical structures which is the key idea in understanding the turbulent flows. This can be used to identify the large energy containing structures resolved by hybrid RANS-LES models to assess the model accuracy as compared to experiment or more accurate computational models (DNS, LES).

Periodic hills geometry has been investigated both numerically and experimentally over a wide range of Reynolds number by several authors. One of the first works was the experimental investigation of \citep{almeida1993wake}. The domain size of Almeida's experiment made it computationally expensive to perform a suitable numerical simulation. There were also concerns if the periodic boundary conditions may be applicable for his experiments. To overcome this issue, \citep{mellen2000large} performed experiments on the periodic hills to provide validation data for simulations. Parallely, computational work using LES was also performed to provide validation data for RANS and hybrid methods \citep{frohlich2005highly, breuer2009flow}. \citep{frohlich2005highly} performed highly resolved LES at Re = 10595, based on hill height. They presented detailed results of mean flow quantities, Reynolds stresses, and budget for the Reynolds stresses. \citep{breuer2009flow} performed combined numerical and experimental work for a range of Reynolds number. They reported results about the existence of small recirculation zone at hill top and size of the recirculation zone. Recently, \citep{diosady2014dns} performed DNS at Re = 10595 using an $8^{th}$ order scheme in space and $4^{th}$ order scheme in time. 

The simplicity of the geometry coupled with complex flow physics has made periodic hills simulations a popular choice for testing hybrid RANS-LES models in the literature. Some of the early works to demonstrate the reduction in computational cost using the hybrid models used periodic hills as a test case \citep{davidson2003hybrid,davidson2005hybrid,temmerman2005hybrid,tessicini2006approximate}. The general observation in all these studies was that the use of hybrid RANS-LES models reduces the computational cost compared to LES while providing better accuracy than RANS. To compare the performance of the different hybrid models, a joint study was performed \citep{vsaric2007evaluation}  by various groups to study the flow over periodic hills at $Re=10595$ and numbers of difference in the location of the separation and reattachment points had been observed.

In this paper, unsteady flow field prediction has been compared for a channel flow with periodic constrictions at a Reynolds number Re=10595 for a set of fine and coarse grids using local dynamic k-equation subgrid-scale LES model \citep{kim1995new}. Also, a comparison has been made with simulation for the coarse grid using a hybrid RANS-LES model for statistics and unsteady flow dynamics. POD has been used to verify the cascading of energy into different POD Eigen-modes. 
  
The rest of the paper is organized as follows. The mathematical models used for the simulations are presented in Sec. 2. Section 3 provides the details of the grid generation, numerical solver and the list of simulations that have been performed. The statistical comparison of results is obtained in Sec. 4.1 and in section 4.2, energy cascading has been compared using POD for LES and Hybrid RANS-LES models. Conclusions are presented in Section 5.

\section{Mathematical Model }
\subsection{Flow Field Equations}
The governing equations for the conservation of mass and momentum for incompressible flows are given by
\begin{equation}
  \frac{\partial \bar{U}_i}{\partial x_i}=0 \label{Eq:continuity}
\end{equation}
\begin{equation}
  \frac{\partial \bar{U}_i}{\partial t}+\frac{\partial \bar{U}_i \bar{U}_j}{\partial x_j}=
  -\frac{\partial \bar{p}}{\partial x_i}+\nu \frac{\partial^2 \bar{U}_i}{\partial x_j \partial x_j}
  +\frac{\partial \tau_{ij}}{\partial x_j} \label{Eq:momentum}
\end{equation}
Here $\mbox{ }\bar{}\mbox{ }$ is used to represent the hybrid variables. It is assumed that the variables seamlessly switch between the ensembled RANS and filtered LES variables.  $\bar{U}_i$ is the fluid velocity, $\bar{p}$ is the fluid pressure, $\nu$ is the kinematic viscosity and $\tau_{ij}$ is the turbulent stress tensor. The turbulent stress tensor is given by
\begin{equation}
  \tau_{ij}=\frac{2}{3}k\delta_{ij}-2\nu_t^h\bar{S}_{ij}+N_{ij} \label{Eq:turbulent_stress_tensor}
\end{equation}
Here $k$ is the turbulent kinetic energy (TKE), $\nu_t^h$ is the hybrid turbulent viscosity, $\bar{S}_{ij}$ is the symmetric part of the velocity gradient tensor  and $N_{ij}$ is the non-linear part of turbulent stress tensor. For linear eddy-viscosity models, $N_{ij}=0$. In this study, the constitutive relation proposed by Abe et. al. \citep{abe2005hra} is used for $N_{ij}$ in hybrid RANS-LES model. The turbulent kinetic energy is calculated using the following equation
\begin{equation}
  \frac{\partial k}{\partial t}+\frac{\partial k \bar{U_j}}{\partial x_j}= -\tau_{ij}\frac{\partial \bar{U}_i}{\partial x_j}
  - \epsilon^h+\frac{\partial}{\partial x_j}\left[\left(\nu+\frac{\nu_t^h}{\sigma_k}\right)\frac
    {\partial k}{\partial x_j}\right] \label{Eq:hybrid_tke}
\end{equation}
The system of equations are closed once the hybrid turbulent viscosity $\nu_t^h$ and dissipation rate $\epsilon^h$ are provided. This is discussed in section \ref{hybridModel}.

\subsection{RANS Model}
In this section details of the baseline RANS models are provided. K-$\omega$ SST model \citep{mentersst} is used as the baseline linear RANS model. The model solves two transport equation for the TKE and specific turbulent dissipation rate $\omega$  \citep{mentersst}
\begin{equation}
	\frac{\partial k}{\partial t}+\bar{u}_{j}\frac{\partial k}{\partial x_{j}}=P_{k}-\beta^{*}\omega k+\frac{\partial}{\partial x_{j}}\left[\left(\nu+\sigma_{k}\nu_{t}\right)\frac{\partial k}{\partial x_{j}}\right]\label{eq:kwsst_k}
\end{equation}
\begin{align}
	\frac{\partial\omega}{\partial t}+\bar{u}_{j}\frac{\partial\omega}{\partial x_{j}}&=\frac{\gamma}{\nu_{t}}P_{k} 
	-\beta\omega^{2}+\frac{\partial}{\partial x_{j}}\left[\left(\nu+\sigma_{w}\nu_{t}\right)\frac{\partial\omega}{\partial x_{j}}\right]  
	\\ \nonumber &\qquad {}+ 2\left(1-F_{1}\right)\frac{\sigma_{w2}}{\omega}\frac{\partial k}{\partial x_{j}}\frac{\partial\omega}{\partial x_{j}}\label{eq:kwsst_w}
\end{align}
where $P_{k}=\max(\tau_{ij}\partial\bar{u}_{i}/\partial x_{j},10\beta^*\omega k$),
is the kinetic energy production term, $\beta^{*}=0.09$ is a model constant
and the last term in the $\omega$ equation is the cross-diffusion
term. $F_{1}$ is a blending function which has a value of one inside
the boundary layer and zero outside.  The limiting of the production term is an alternative to the use of damping function in the near-wall region. The turbulent stress tensor and viscosity are computed
in this model as follows
\begin{equation}
	\tau_{ij}=\frac{2}{3}k\delta_{ij}-2\nu_t \bar{S}_{ij}+N_{ij}
	\label{eq:kwsst_stress}
\end{equation}
\begin{equation}
	\nu_{t}=\frac{a_{1}k}{\max\left(a_{1}\omega,\sqrt{2\bar{S}_{ij}\bar{S}_{ij}}F_{2}\right)}\label{eq:kwsst_turbviscosity}
\end{equation}
Here $N_{ij}$ is the non-linear part of the turbulent stress tensor. For more details on k-$\omega$ SST RANS models and non linear part of turbulent stress tensor, see Appendices \ref{appendix}.

\subsection{LES Model}
In this study, local dynamic k-Equation Subgrid-Scale (LDKSGS) Model\citep{kim1995new} is used for LES computations. Here, subgrid stresses $\tau_{ij}$ are modelled in terms of the SGS eddy viscosity $\nu_t$ as: 
\begin{equation}
\tau_{ij} = -2\nu_t\widetilde{S_{ij}} + \frac{2}{3}\delta_{ij}k_{sgs}
\end{equation}
where, $\mbox{ }\widetilde{ }\mbox{ }$ represents filtering, filter width is taken to be the cube-root of the cell volume and 
\begin{equation}
\nu_{sgs} = c_\nu k^\frac{1}{2}_{sgs}\Delta
\end{equation}A one-equation model for the subgrid-scale kinetic energy, $k_{sgs}$ is given in the following form:
\begin{equation}
  \frac{\partial k_{sgs}}{\partial t}+\frac{\partial k_{sgs} \widetilde{U_j}}{\partial x_j}= -\tau_{ij}\frac{\partial \widetilde{U}_i}{\partial x_j}
  - \epsilon+\frac{\partial}{\partial x_j}\left(\nu_{sgs}\frac
    {\partial k}{\partial x_j}\right) \label{Eq:LES_tke}
\end{equation}
Equation \ref{Eq:LES_tke} is closed by providing a model for dissipation rate term $\epsilon$ as:
\begin{equation}
\epsilon = c_\epsilon \frac{k^\frac{3}{2}_{sgs}}{\Delta}
\end{equation}
Here two new coefficients $c_\nu$ and $c_\epsilon$ are calculated dynamically using the method proposed in \citep{kim1995new} along with local averaging of model coefficients for stability.
\subsection{Hybrid Model} \label{hybridModel}
A general formulation for the viscosity and dissipation in hybrid methods can be written as follows 
\begin{equation}
\epsilon^h=F_\epsilon(\epsilon^r,\epsilon^l)=k^{3/2}F_\epsilon(L_\epsilon^r,L_\epsilon^l) \label{Eq:hybrid_dissipation}
\end{equation}
\begin{equation}
\nu_t^h=F_\nu(\nu_t^r,\nu_t^l)=\sqrt{k}F_\nu(L_\nu^r,L_\nu^l) \label{Eq:hybrid_viscosity}
\end{equation}
In these equations, superscripts $h$, $r$, and $l$ denotes hybrid, RANS and LES, respectively. $L_\nu$ and $L_\epsilon$ denote the turbulent viscosity and dissipation length scales. Both these scales do not have to be the same \citep{breuer2008hrt}. $F_\epsilon$ and $F_\mu$ denote the hybrid RANS-LES switching function based on the length scales. It is also possible to formulate the models in terms of the turbulent time-scales. In this study, we propose to investigate the performance of SST-blended hybrid RANS-LES model which uses hyperbolic tangent switching function for both dissipation and viscosity. The expressions for the length scales and switching functions for the hybrid model are given in Table \ref{Tab:hybrid_models}.

{\renewcommand{\arraystretch}{1.5} 
\begin{table}
  \centering%
  \protect\caption{In the expressions given, $\beta^*$ is model constant. $F_b$ is the blending function.}
\begin{tabular}{|c|c|c|c|c|c|c|}
\hline 
Model & $L_\epsilon^r$  & $L_\epsilon^l$ & $L_\nu^r$ & $L_\nu^l$ & $F_\epsilon$ & $F_\nu$  \tabularnewline
\hline
SST-Blended &  $\frac{\sqrt{k}}{\beta^* \omega}$ & $\Delta$ &  $\frac{\sqrt{k}}{\beta^* \omega}$ & $\Delta$ & $\frac{1}{F_bL_\epsilon^r+(1-F_b)L_\epsilon^l}$ & $F_bL_\nu^r+(1-F_b)l_\nu^l$ \tabularnewline
\hline
\end{tabular}
\label{Tab:hybrid_models}
\end{table}
}
In the blended model, both $\nu_t^h$ and $\epsilon^h$ are modified from their RANS values. The switching occurs over a number of grid cells in the blended model (also called buffer region). Blended model uses complex switching function. 
\begin{equation}
  F_b=1-0.5\left[1+\tanh\left(\frac{1-L_\nu^l/L_\nu^r}{\lambda}\right)\right] \label{Eq:BSST_blending_funcion}
\end{equation}

The amount of blending in SST-Blended is controlled by the model constant $\lambda$. This parameter is set to 0.25 in the current study \citep{gopalan_cost}.

The filter width is taken to be the square-root of the maximum face area of the cell. Characteristic length scale of turbulence is used for $L_\nu^r$ and $L_\epsilon^r$ in the two-equation model. The shear stress transport (SST) model of Menter \citep{mentersst} is adopted as the baseline two-equation RANS model for hybrid model. Non-linear hybrid model is denoted by adding prefix ``N'' to the linear models (eg. NBSST). The hybrid model includes shielding function to avoid modelled stress depletion and grid induced separation. The model uses the SST blending function $F_2$ (Eq. \ref{eq:kwsst_F2}) as the shielding function (see Appendices).

The definition of a generic framework for hybrid turbulent viscosity and dissipation (Eqs. \ref{Eq:hybrid_dissipation}-\ref{Eq:hybrid_viscosity}) allows the construction of a model free hybrid approach and makes it easier to modify existing RANS codes to create the new hybrid models. 

\subsection{Proper Orthogonal Decomposition}
This method is adopted from the formulation presented in \citep{berkooz1993proper}. In the current analysis, POD is applied on real valued scalar and vector fields and a simplified mathematical formulation is described as follows.

Consider a real valued scalar field f defined on an interval $\Omega$. The inner product (f,g) and norm $||f||$ defined as 
\begin{equation}
\begin{split}
(f,g) &= \int_\Omega f(x)g(x)dx \\
 &= <f*g> \text{where $<.>$ means ensemble}
\end{split}
\end{equation}
\begin{equation}
||f|| = (f*f)^\frac{1}{2}
\end{equation}

The problem of finding a single deterministic function most similar, on an average, to the set of values observed for the function u(x) mathematically translates to seeking a function $\phi(x)$ such that 
\begin{equation}
max\Bigg( \frac{<||(u,\psi)||^2>}{(\psi,\psi)}\Bigg)_\psi = \frac{<||(u,\phi)||^2>}{(\phi,\phi)}
\label{POD_psi}
\end{equation}
That is, we find the member of the $\psi(=\phi)$ which maximises the normalised inner product with the field u, which is most nearly parallel in function space. A necessary condition for (\ref{POD_psi}) to hold is that $\phi$ is an eigen-function of the two-point auto-covariance matrix $R(x,x') = <u(x)u(x')>$. 
\begin{equation}
\int_{\Omega}<u(x)u(x')>\phi(x')dx' = \lambda\phi(x)
\label{POD_phi}
\end{equation}

The maximum in (\ref{POD_psi}) is obtained for largest eigenvalue $\lambda_1$ of (\ref{POD_phi}), However, Hilbert Schmidth theory assures that there is denumerable infinity of solutions of (\ref{POD_phi}) as long as $\Omega$ is bounded. These are called empirical eigen-functions and we denote these by $\{\phi_k\}$ and normalise them so that $||\phi_k||$ = 1. We order the eigenvalues by $\lambda_k > \lambda_{k+1}$. Observing the non-negative definiteness of R(x,x') assures that $\lambda_k > 0$. Also, the ensemble may be reproduced by a modal decomposition in the eigen-functions:
\begin{equation}
u(x) = \Sigma_ka_k\phi_k(x)
\end{equation}
The diagonal decomposition of the two-point auto-covariance matrix R ensures that the modal amplitudes are uncorrelated:
\begin{equation}
\begin{split}
R(x,x') &= \Sigma_k\lambda_k\phi_k(x)\phi_k(x')\\
<a_k a_{k'}> &= \delta_{kk'}\lambda_k
\end{split}
\end{equation} 
This method can be used to reduce a dynamical system with infinite degrees of freedom represented by infinite number of eigen-modes, possible to be obtained using equation (\ref{POD_phi}), to a system with finite number of modes (N) which is most similar to the original system on an average. This finite number of modes N will depend on the smallness of eigen-values of the higher modes and could be seen as a loss of the information about the system. This reduction of the dynamical system to a finite degrees of freedom also provides a finite number of empirical eigen-function which represents the dynamics of the POD modes obtained and can be used to understand the evolution of the system considered. 

\section{Numerical Details }
Computational setup for the simulations is shown in Fig. \ref{fig:domain}. The size of the domain is defined in terms of the height of the hill H. The domain size is taken to be 9H$\times$3.035H$\times$4.5H in streamwise (x), wall-normal (y)  and spanwise (z) directions, respectively. The size has been chosen to match the benchmark LES results from the literature \citep{breuer2009flow}.

\subsection{Grid Generation}
\begin{figure}
	\centering
	\begin{minipage}{150mm}
		\subfigure[Domain setup]{
			\resizebox*{7cm}{!}{\includegraphics[width=\textwidth]{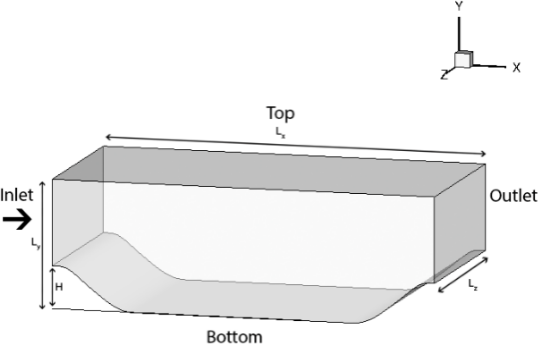}}
			\label{fig:domain}
		}
		\subfigure[Grid used for the computations]{
			\resizebox*{7cm}{!}{\includegraphics[width=\textwidth]{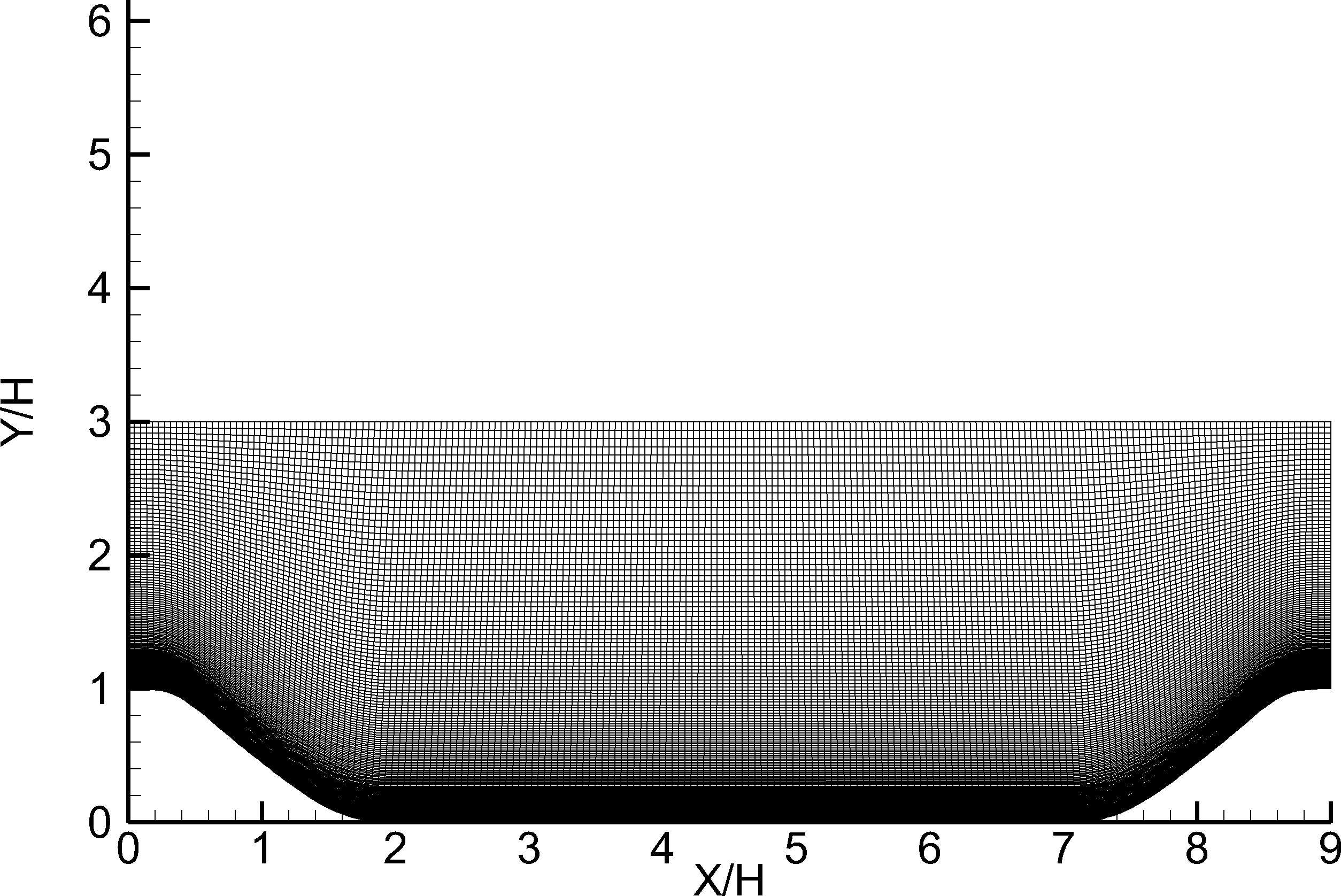}}
			\label{fig:grid_cross_section}}
		\caption[]{Computational setup for the simulations}
	\end{minipage}
\end{figure}

Grids used in the current study are generated using commercial meshing software ANSYS-ICEMCFD$^\circledR$. Figure \ref{fig:grid_cross_section} shows a cross-section (x-y) of the grid. Five different grids are generated and the details of the grids are given in Table \ref{tab:grid_details}. The grids G1, G2 and G3 are used to perform Hybrid RANS-LES simulations and grids G2, G3, G4 and G5 are used to compute LES simulations. The average value of $y^+$ is found to be in the range 0.2-0.3 for all the grids. 

\subsection{Numerical Solver}
All the simulations are performed using the open-source CFD toolbox OpenFOAM$^\circledR$. The hybrid models have been implemented and linked as a user defined library at run-time. Pressure-velocity coupling is achieved using the PIMPLE algorithm available in the code, which is a blend of the PISO and SIMPLE algorithm. The convection term in the momentum equation is discretized using second-order central difference scheme and for the turbulent variables using bounded second-order schemes. Time marching is performed using a second-order backward difference scheme. All other terms are discretized using central difference schemes. The tolerance has been set to $10^{-6}$ for all the variables. Once a statistically steady state is achieved, time averaging is performed over 20 flow-through times ($L_x/U_b$) to compute the statistics. Periodic boundary conditions are used in streamwise direction  and spanwise directions. No-slip boundary conditions are enforced at the top and bottom wall. As the value of $y^+$ lies in the viscous region, no wall-function was employed and the turbulence variables are directly integrated up to the wall. 
\begin{table}
	\centering
	\caption{Grids used for the simulations.}
	{
		\begin{tabular}{|l l l l l l l|} \hline
			Grid & $n_x$ & $n_y$ & $n_z$ & $\Delta x/H$ & $\Delta y/H$ & $\Delta z/H$ \\ \hline 
			G1 & 127 & 94 & 40 & 0.058 - 0.078 & 0.002 - 0.117 & 0.113\\
			G2 & 159 & 117 & 50 & 0.046 - 0.062 & 0.002 - 0.087 & 0.09\\
			G3 & 200 & 146 & 63 & 0.036 - 0.049 & 0.002 - 0.065 & 0.071\\
			G4 & 200 & 146 & 186 & 0.036 - 0.049 & 0.002 - 0.065 & 0.024\\
			G5 & 252 & 184 & 235 & 0.036 & 0.002 - 0.031 & 0.019\\ \hline
		\end{tabular}
	}
	\label{tab:grid_details}
\end{table}

\section{Results}
\subsection{Grid Comparison}
\begin{figure}
	\centering
	\begin{minipage}{150mm}
		\subfigure[Skin-friction coefficient]{
			\resizebox*{7cm}{!}{\includegraphics[width=\textwidth]{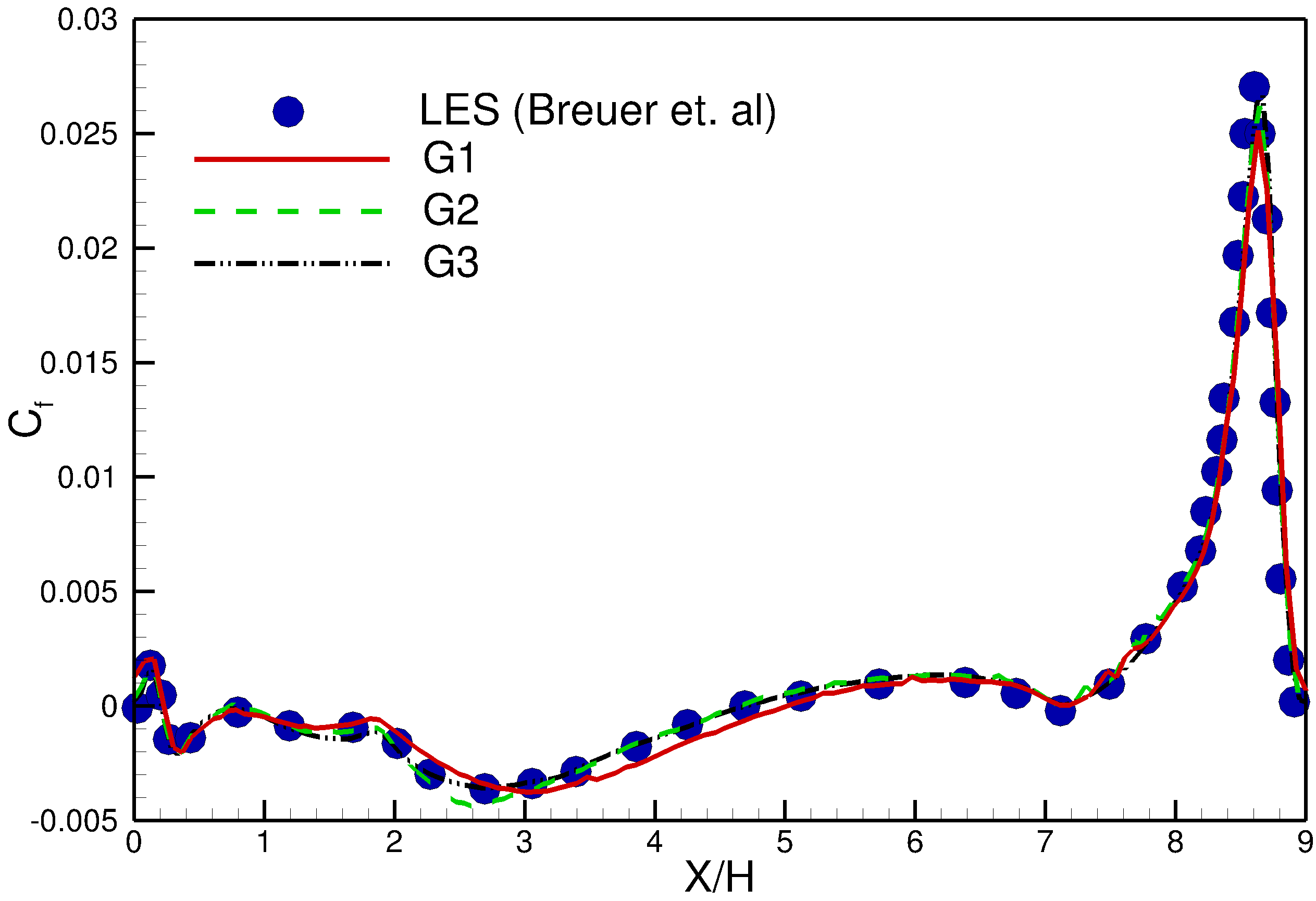}}
			\label{fig:NBSSTCf}
		}
		\subfigure[Wall pressure coefficient]{
			\resizebox*{7cm}{!}{\includegraphics[width=\textwidth]{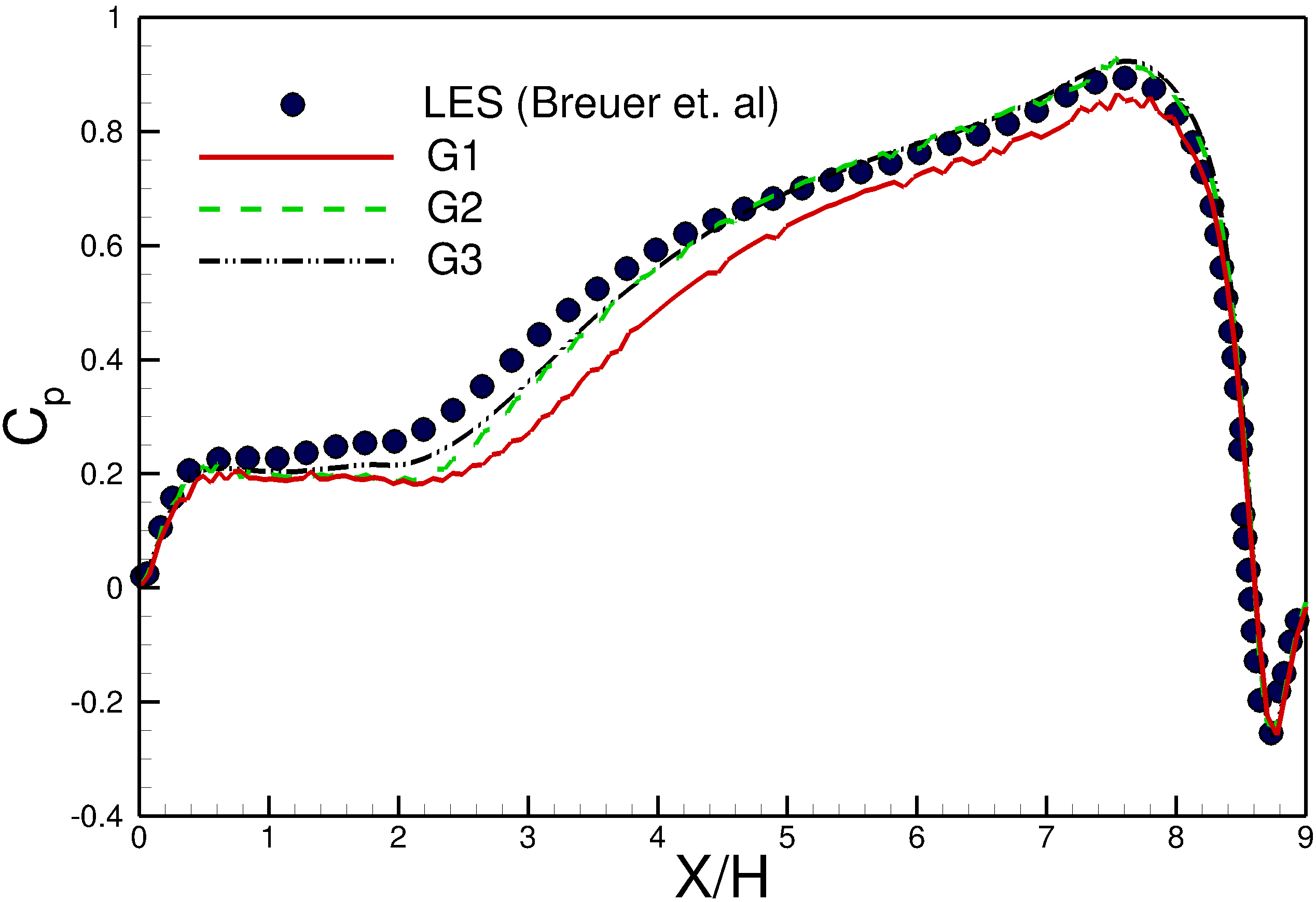}}
			\label{fig:NBSSTCp}}

		\subfigure[Difference in skin-friction coefficient]{
			\resizebox*{7cm}{!}{\includegraphics[width=\textwidth]{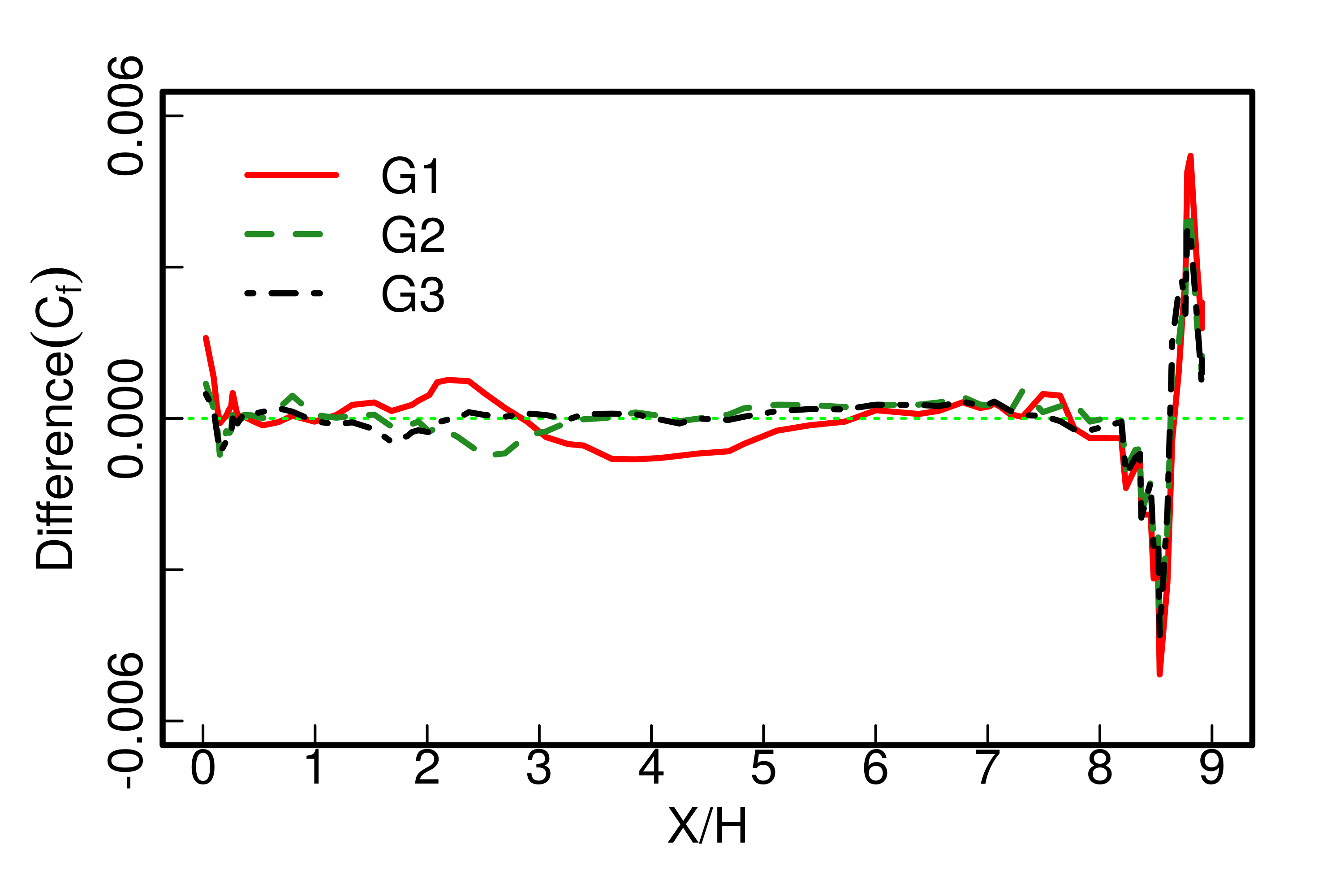}}
			\label{fig:NBSSTCferror}
		}
		\subfigure[Difference in wall pressure coefficient]{
			\resizebox*{7cm}{!}{\includegraphics[width=\textwidth]{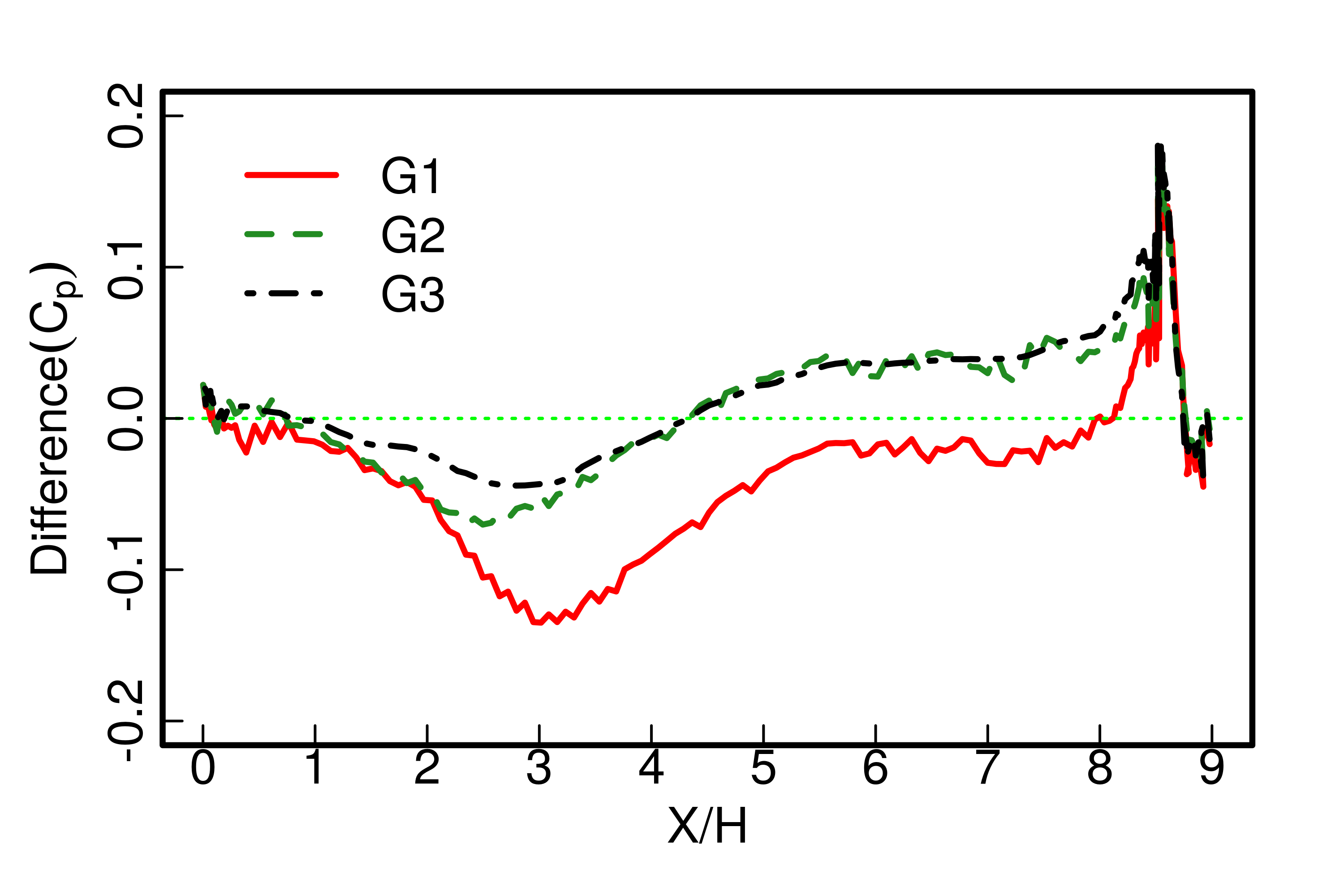}}
			\label{fig:NBSSTCperror}}
		\caption[]{Comparison of the Skin-friction and wall pressure coefficient among G1, G2 and G3 grid with NBSST and previous LES data from Breuer et al.\citep{breuer2009flow}. Difference is calculated with respect to the LES data from Breuer et al.}
		\label{fig:NBSSTwall}
	\end{minipage}
\end{figure}

Figures \ref{fig:NBSSTCf} and \ref{fig:NBSSTCp} show the comparison of the skin-friction and wall pressure coefficients along the channel for the NBSST simulations on G1, G2 and G3 grids and Figures \ref{fig:NBSSTCferror} and \ref{fig:NBSSTCperror} depict the comparison of difference in skin-friction and wall pressure coefficients with respect to data provided in \citep{breuer2009flow}. From these figures, it can be clearly seen that G2 and G3 grid have very close prediction of skin-friction and wall pressure coefficients except minor difference in the middle of the recirculation region near X/H=2.5. Also, G2 and G3 grids show significant improvement in predictions as compared to G1 grid. Hence, G3 grid can be considered as grid with optimum resolution for Hybrid RANS-LES simulation and only G3 grid is chosen for the detailed analysis.

\begin{figure}
	\centering
	\begin{minipage}{150mm}
		\subfigure[Skin-friction coefficient]{
			\resizebox*{7cm}{!}{\includegraphics[width=\textwidth]{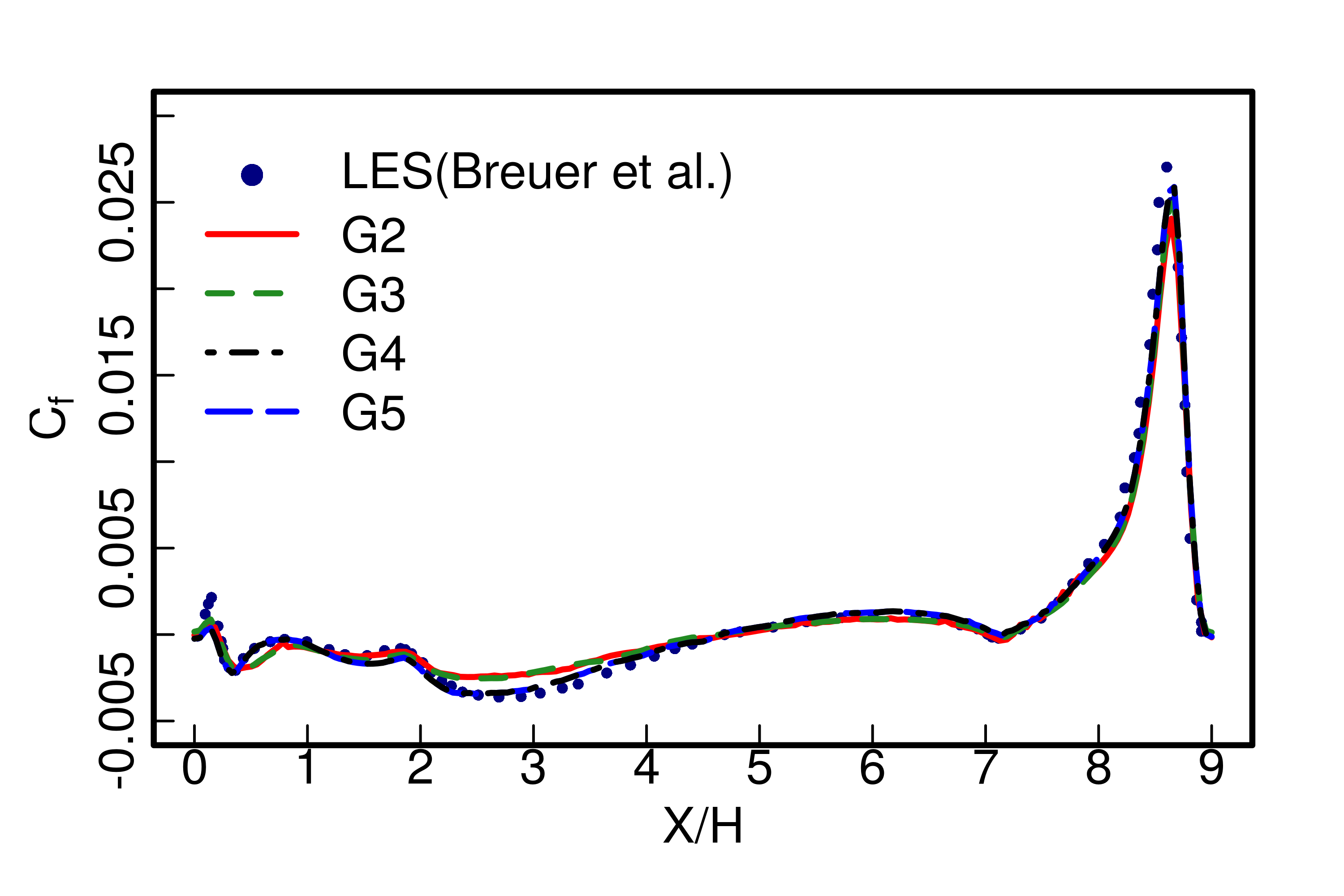}}
			\label{fig:LESCf}
		}
		\subfigure[Wall pressure coefficient]{
			\resizebox*{7cm}{!}{\includegraphics[width=\textwidth]{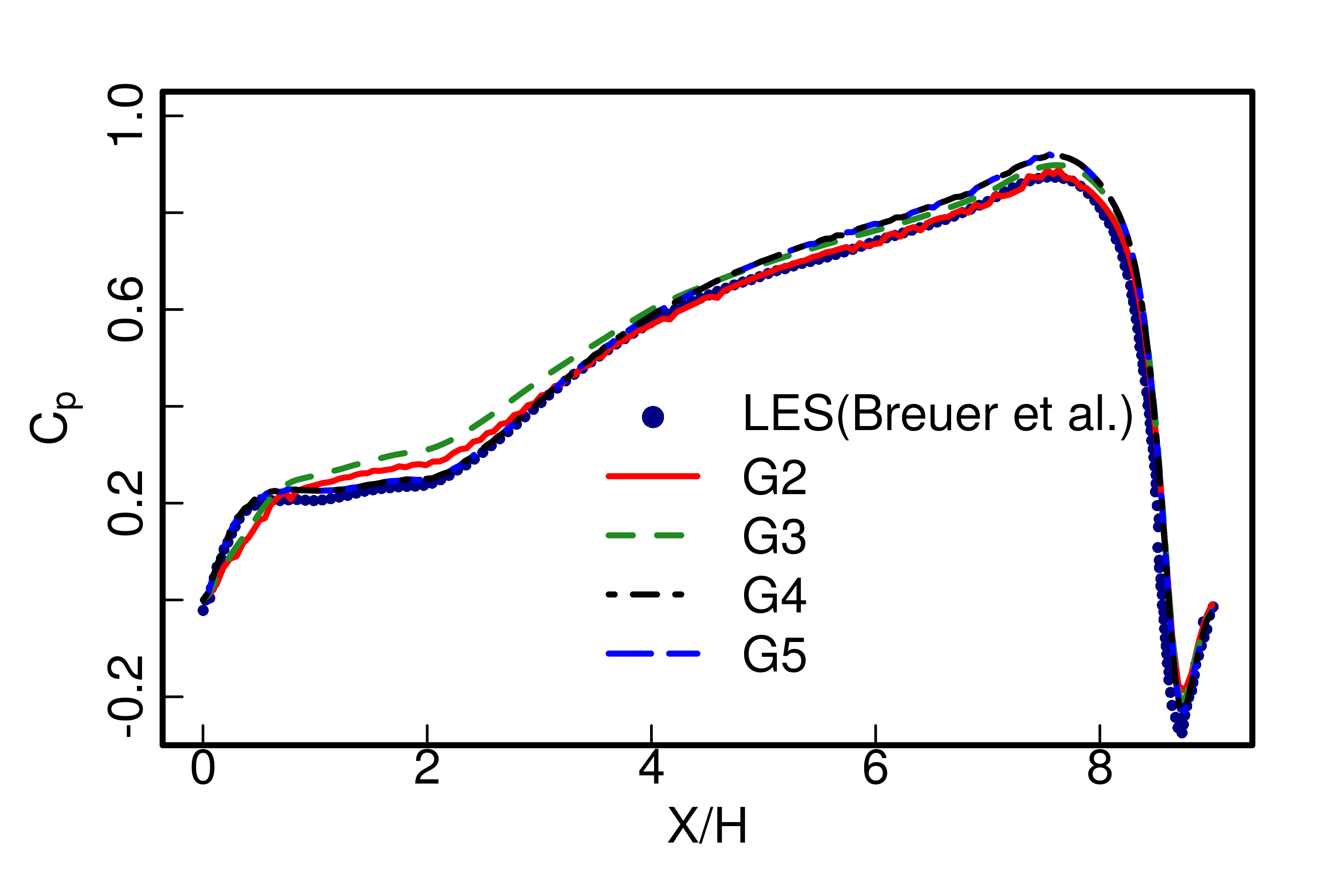}}
			\label{fig:LESCp}}

		\subfigure[Difference in skin-friction coefficient]{
			\resizebox*{7cm}{!}{\includegraphics[width=\textwidth]{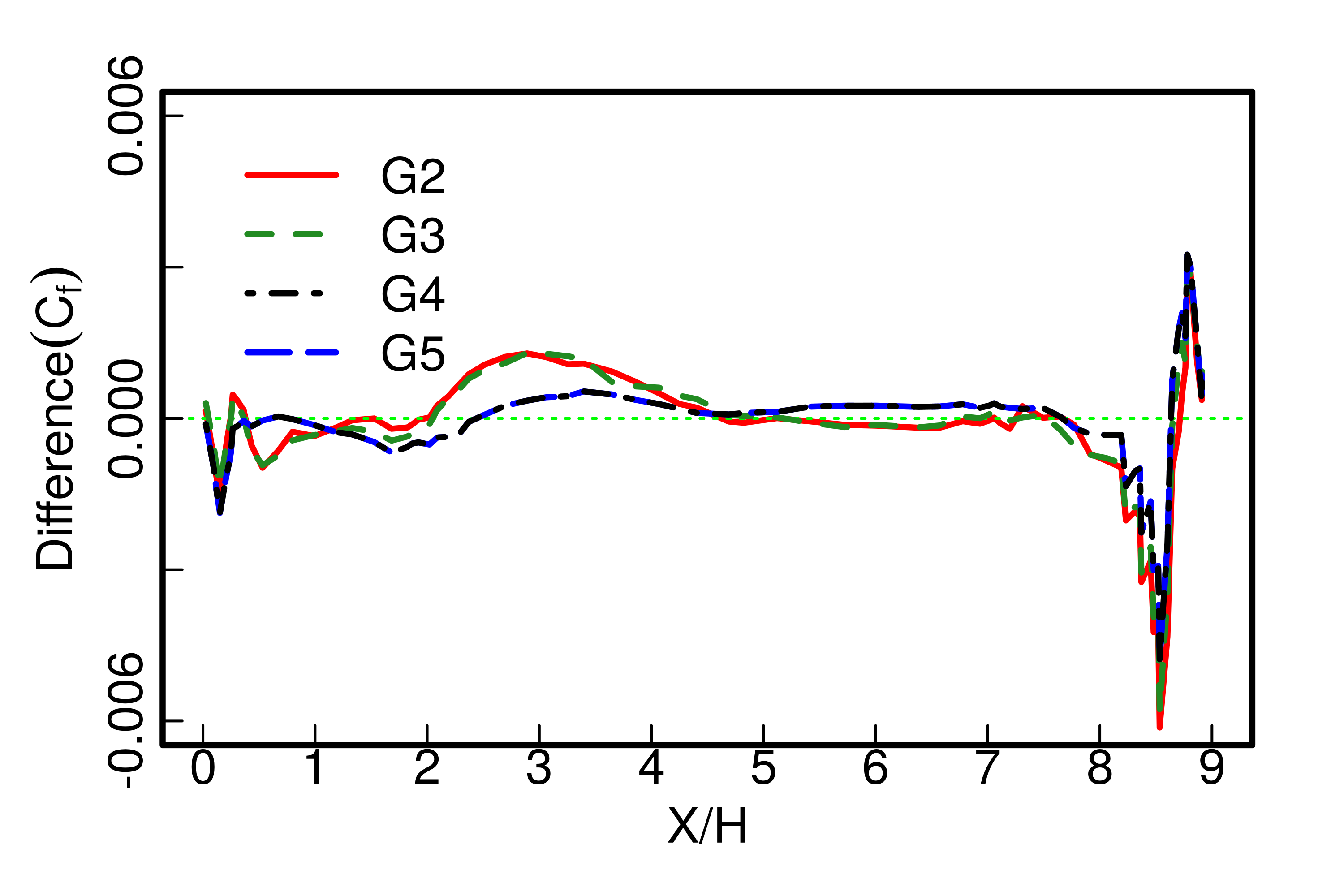}}
			\label{fig:LESCferror}
		}
		\subfigure[Difference in wall pressure coefficient]{
			\resizebox*{7cm}{!}{\includegraphics[width=\textwidth]{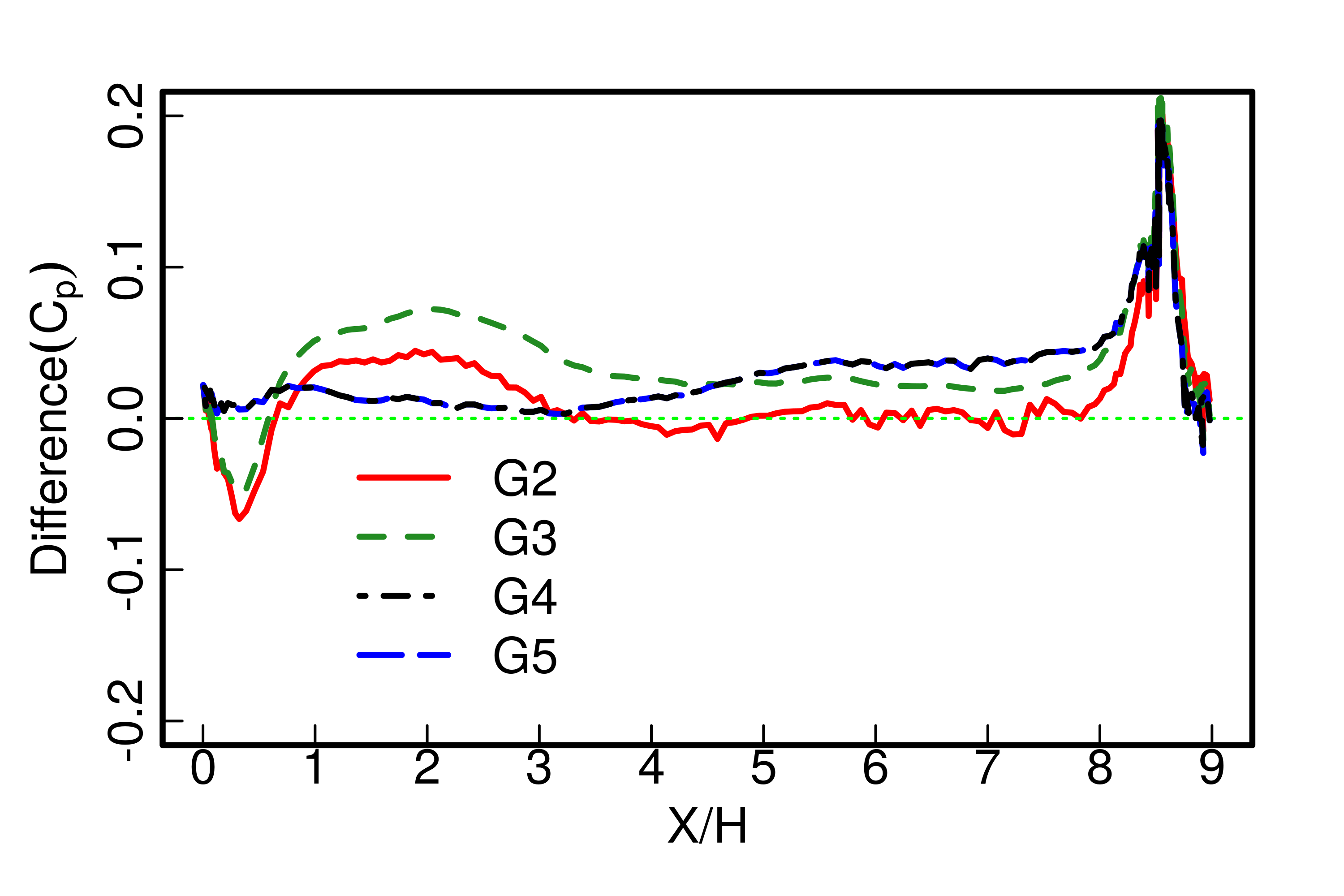}}
			\label{fig:LESCperror}}
		\caption[]{Comparison of the Skin-friction and wall pressure coefficient among G2, G3, G4 and G5 grid with Dynamic k-SGS model and previous LES data from Breuer et. al.\citep{breuer2009flow}. Difference is calculated with respect to the LES data from Breuer et al.}
		\label{fig:LESwall}
	\end{minipage}
\end{figure}

Similarly, grid comparison has been performed on G2, G3, G4 and G5 grids from Table \ref{tab:grid_details} for LES simulations. Figures \ref{fig:LESCf} and \ref{fig:LESCp} show the comparison of the skin-friction and wall pressure coefficients along the hill for the LES on G2 - G5 grids and Figures \ref{fig:LESCferror} and \ref{fig:LESCperror} depict the comparison of difference in skin-friction and wall pressure coefficients with respect to data provided in \citep{breuer2009flow}. Comparing G2 and G3 and G4 and G5, it is evident that increasing grid resolution in streamwise and wall-normal directions does not result in much improvement in predictions of skin-friction and wall pressure coefficients as grid is resolved to $y^+$\textless 1 in the near wall region. However, a significant improvement is observed with spanwise direction grid resolution among G3 and G4. The slight deviation in the skin-friction and wall pressure coefficients prediction in LES simulation on G4 grid and predictions of Breuer et al. may be partly attributed to the use of smagorinsky based LES computation in \citep{breuer2009flow}, whereas Dynamic one-equation subgrid scale model is used here and partly to small difference in grid resolution. Since, G4 grid predictions appear to be very close to the earlier results of \cite{breuer2009flow} and G5 grid does not exhibit much improvement in the results, G4 grid is considered to be optimum for LES calculations and used for further comparison and analysis of results. 

In Table \ref{tab:cost}, the separation and reattachment points for all the simulations along with the computational cost have been tabulated. Location of separation and reattachment point converges to the values reported in \citep{breuer2009flow} using G3 grid for NBSST simulations and G4 and G5 grids for LES simulations. Computational cost for Hybrid RANS-LES simulations on G3 grid is much smaller than LES simulation on G4 grid which provides similar prediction of skin-friction and wall pressure coefficients.  

\begin{table}
	\centering
	\caption{Separation and reattachment points prediction for simulations and computational cost.}
	{
		\begin{tabular}{|l l l l l l|} \hline
			Grid & Model & $X_s/H$ & $X_r/H$ & Grid size & Computational Time for $10^{-3}$ sec. \\  & & & & & physical time with 1 processor\\  \hline
			LES & (\cite{breuer2009flow}) & 0.19 & 4.69 & 12.4$\times10^6$ & - \\ \hline
			G1 & NBSST & 0.23 & 5.0 & 0.48$\times10^6$ & 1119 seconds \\
			G2 & NBSST & 0.22 & 4.65 & 0.93$\times10^6$ & 1526 seconds \\
			G3 & NBSST & 0.21 & 4.7 & 1.84$\times10^6$ & 4450 seconds \\
			G2 & LES & 0.2 & 4.73 & 0.93$\times10^6$ & 1256 seconds \\
			G3 & LES & 0.2 & 4.6 & 1.84$\times10^6$  & 1590 seconds \\
			G4 & LES & 0.17 & 4.7 & 5.43$\times10^6$ & 7488 seconds \\
			G5 & LES & 0.17 & 4.7 & 10.9$\times10^6$ & 144800 seconds \\ \hline
		\end{tabular}
	}
	\label{tab:cost}
\end{table}
\subsection{Mean Statistics}
\begin{figure}
	\centering
	\begin{minipage}{150mm}
		\subfigure[Streamwise velocity]{
			\resizebox*{7cm}{!}{\includegraphics[width=\textwidth]{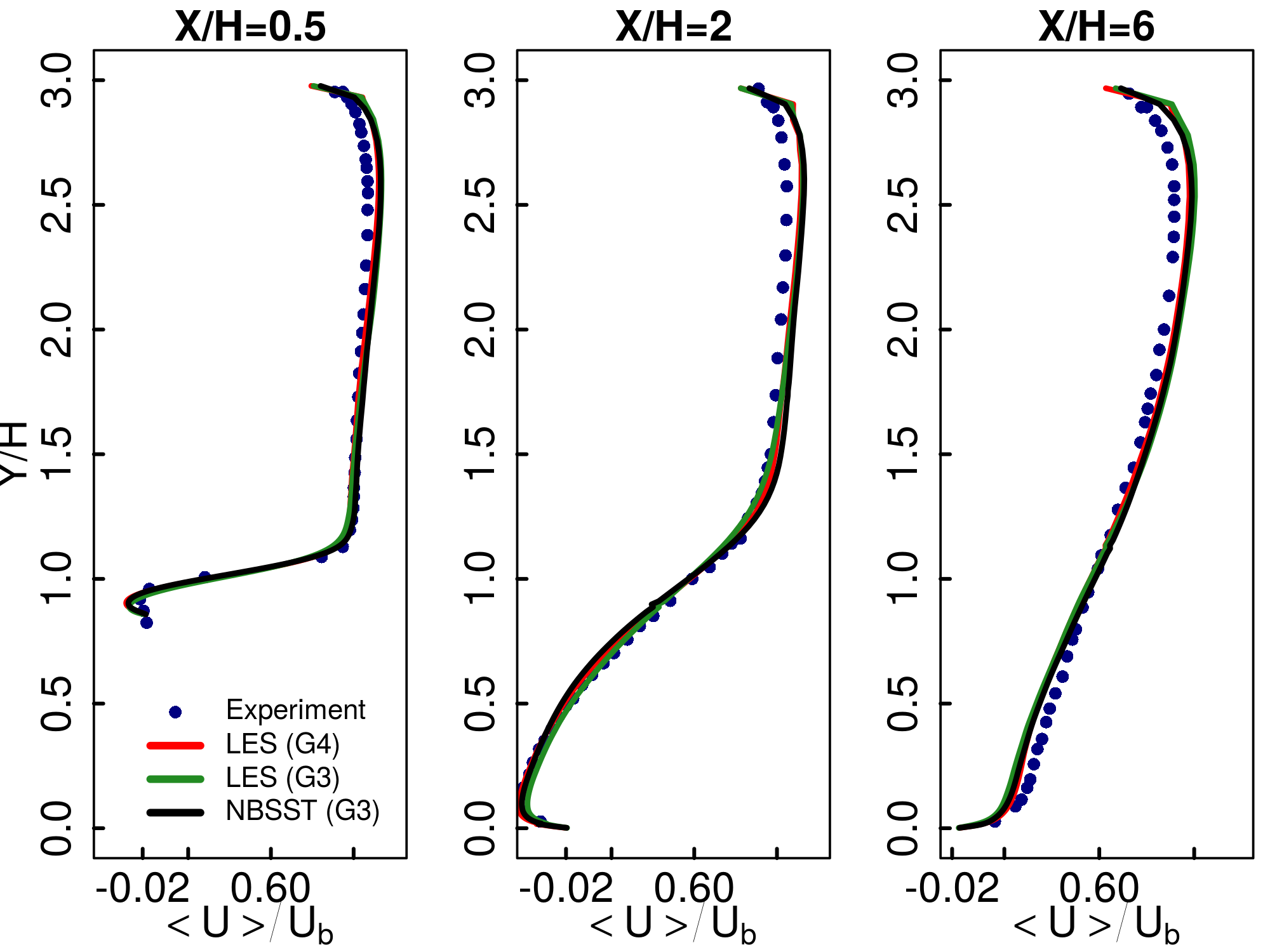}}
			\label{fig:LESU}
		}
		\subfigure[Wall-normal velocity]{
			\resizebox*{7cm}{!}{\includegraphics[width=\textwidth]{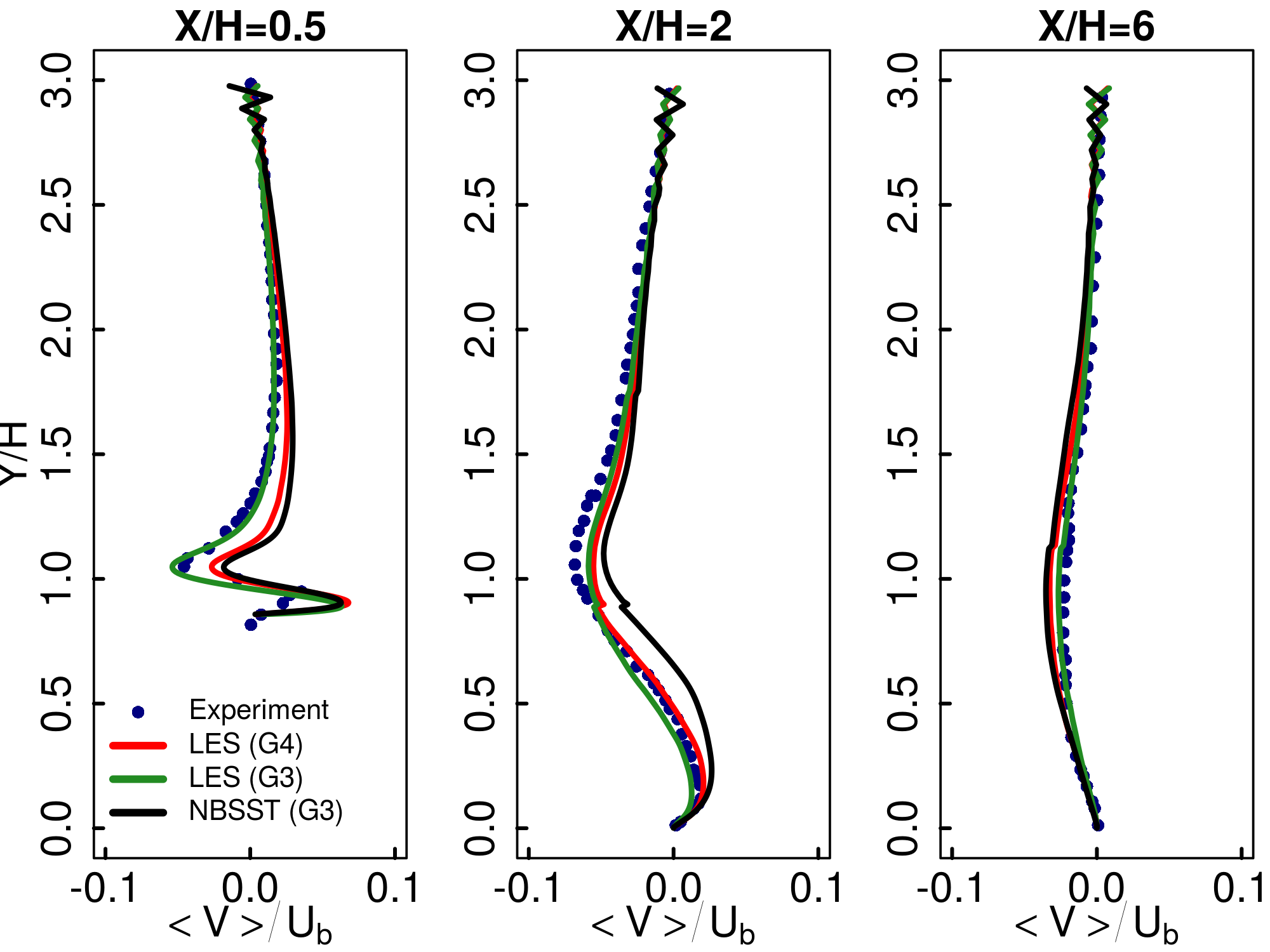}}
			\label{fig:LESV}}

		\subfigure[Mean streamwise Reynolds normal stress]{
			\resizebox*{7cm}{!}{\includegraphics[width=\textwidth]{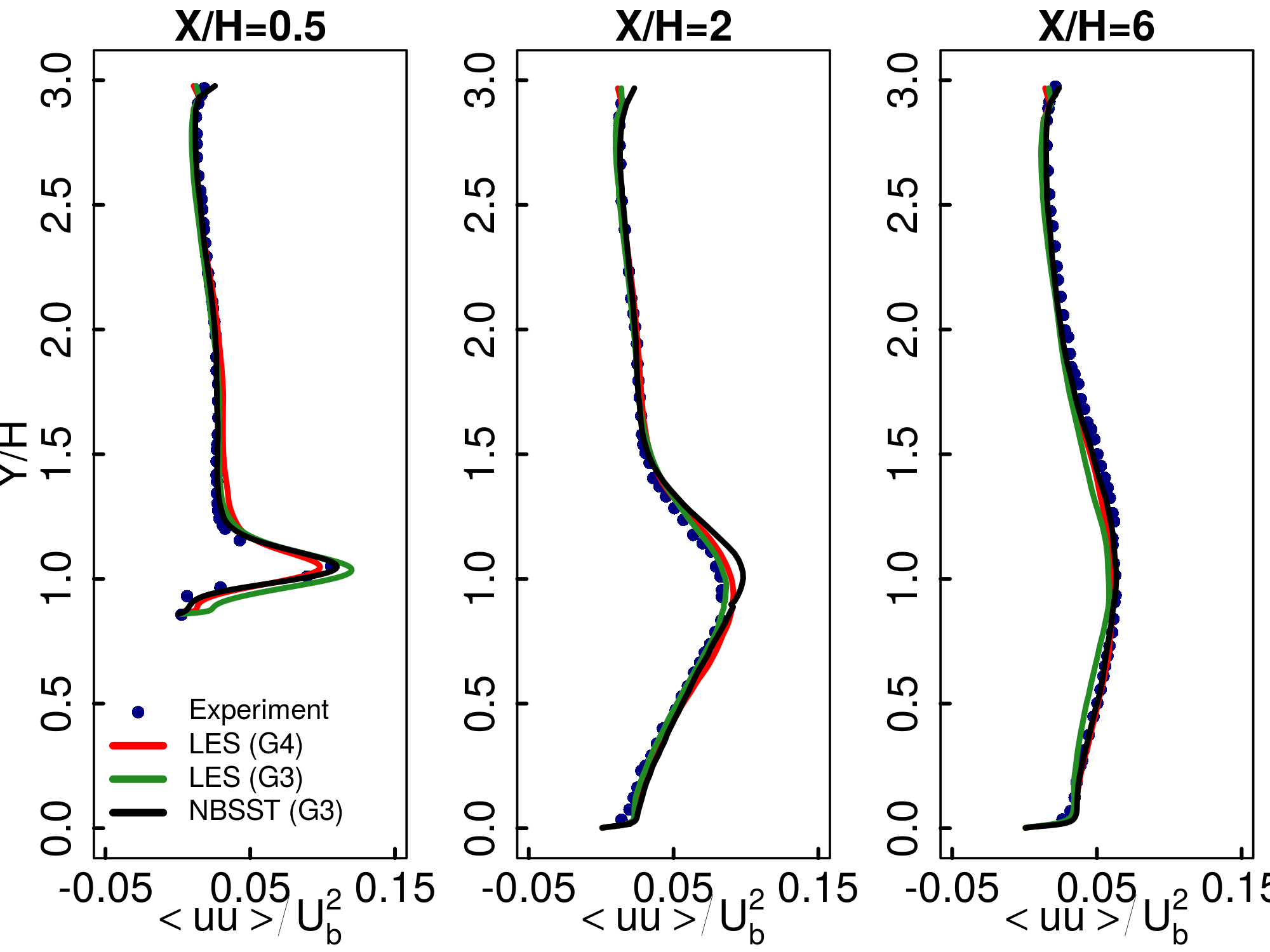}}
			\label{fig:LESuu}
		}
		\subfigure[Mean wall-normal Reynolds normal stress]{
			\resizebox*{7cm}{!}{\includegraphics[width=\textwidth]{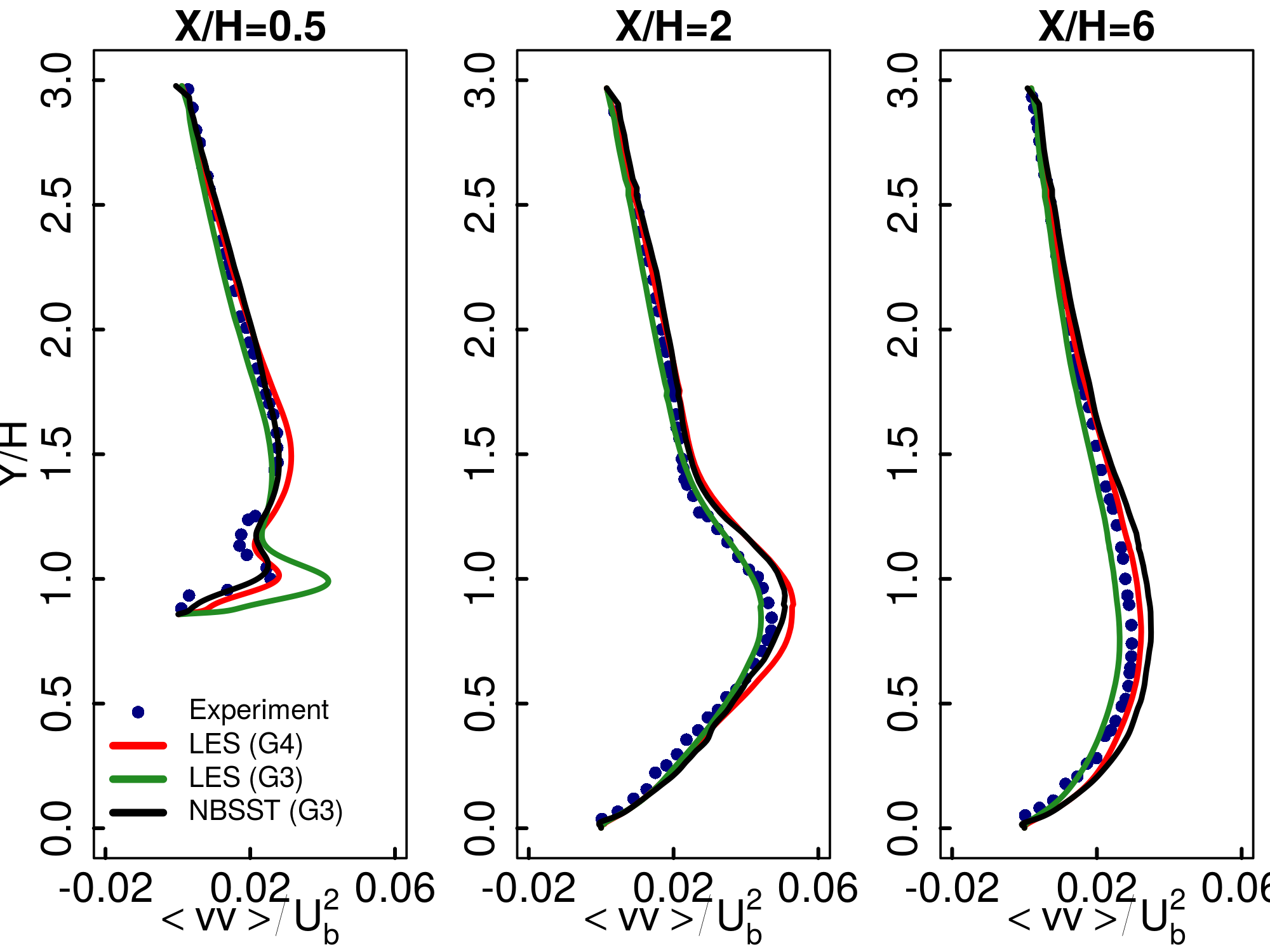}}
			\label{fig:LESvv}}
		
		\subfigure[Mean Reynolds shear stress]{
			\resizebox*{7cm}{!}{\includegraphics[width=\textwidth]{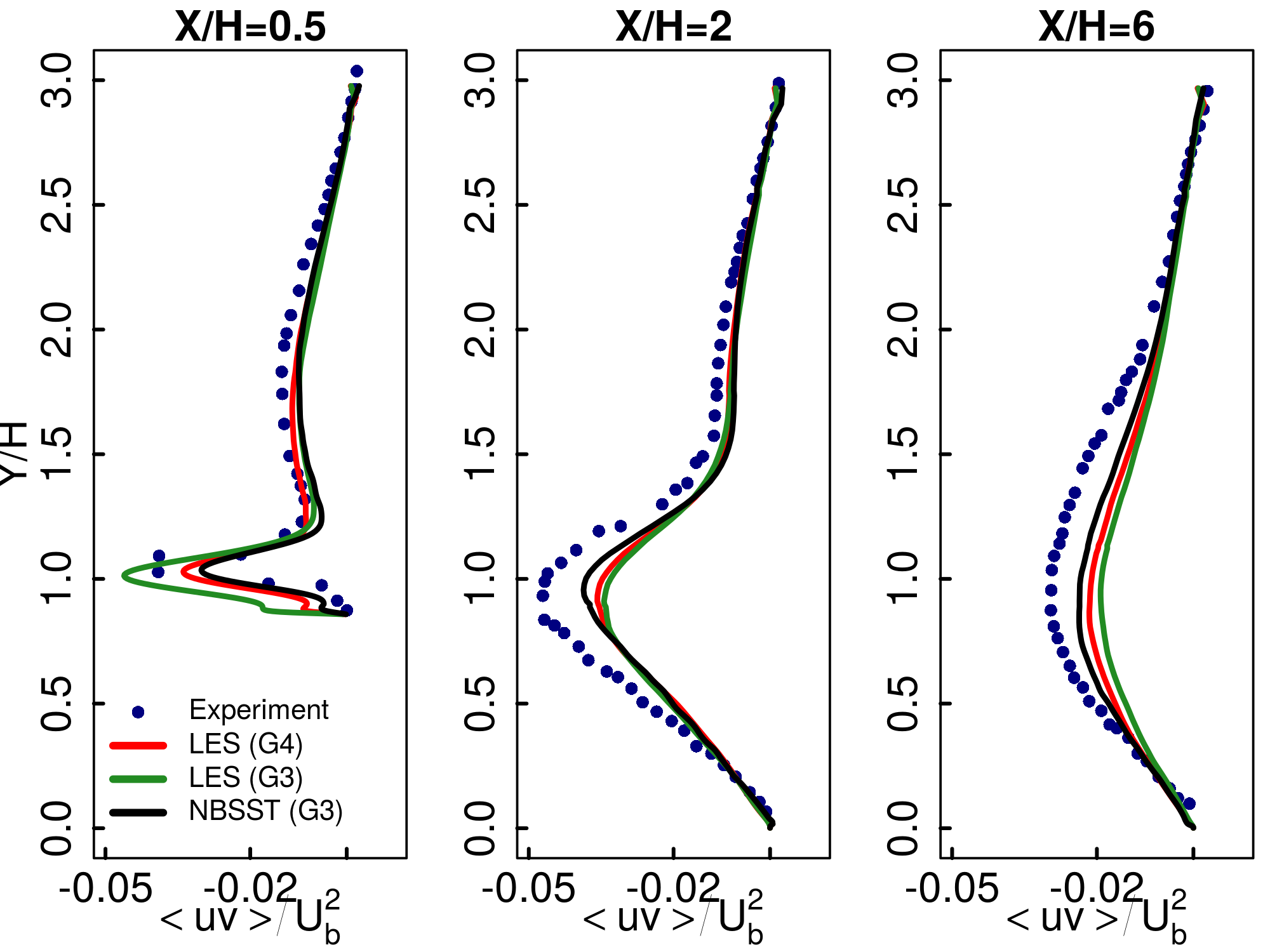}}
			\label{fig:LESuv}}
		\subfigure[Mean turbulent kinetic energy]{
			\resizebox*{7cm}{!}{\includegraphics[width=\textwidth]{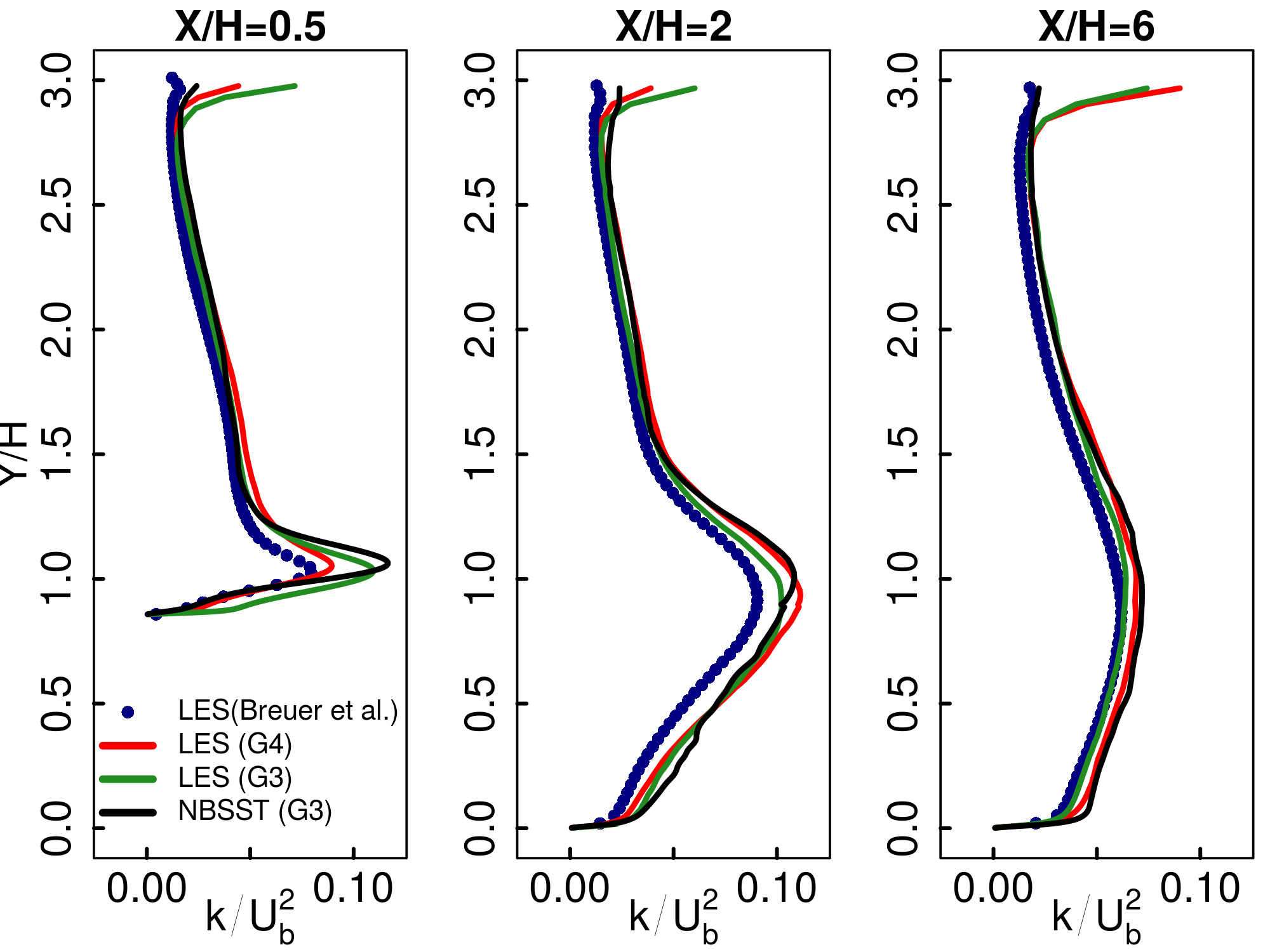}}
			\label{fig:LESk}}
	\end{minipage}
	\caption{Mean flow statistics}
	\label{mean}
\end{figure}

Figures \ref{fig:LESU} and \ref{fig:LESV} depict the comparison of streamwise and wall-normal mean velocity components from LES simulation on G4 grid and NBSST simulation on G3 grid with the experimental results. Difference in streamwise mean velocities among the two simulations is indistinguishable, but LES simulation on G3 grid shows under-prediction of wall-normal velocity at X/H=0.5. Also, NBSST simulation on G3 grid shows slight over-prediction of wall-normal velocity at X/H=2.

Figures \ref{fig:LESuu}, \ref{fig:LESvv}, \ref{fig:LESuv} and \ref{fig:LESk} show comparison of mean streamwise and wall-normal Reynolds normal stress, mean Reynolds shear stress and mean turbulent kinetic energy with experiment. All the predictions by NBSST simulation on G3 grid is in close agreement with experiment and LES simulation on G4 grid, whereas LES simulation on G3 grid shows overprediction of Reynolds normal and shear stresses at X/H=0.5 in the vicinity of the wall. It should be noted here that the correct Reynolds stress components prediction is only possible here due to the use of non-linear corrections in the hybrid RANS-LES model, otherwise a significant underprediction will be observed for linear Hybrid RANS-LES models.  

\begin{figure}
	\centering
	\begin{minipage}{150mm}
		\subfigure[LES (G4)]{
			\resizebox*{7cm}{!}{\includegraphics[width=\textwidth]{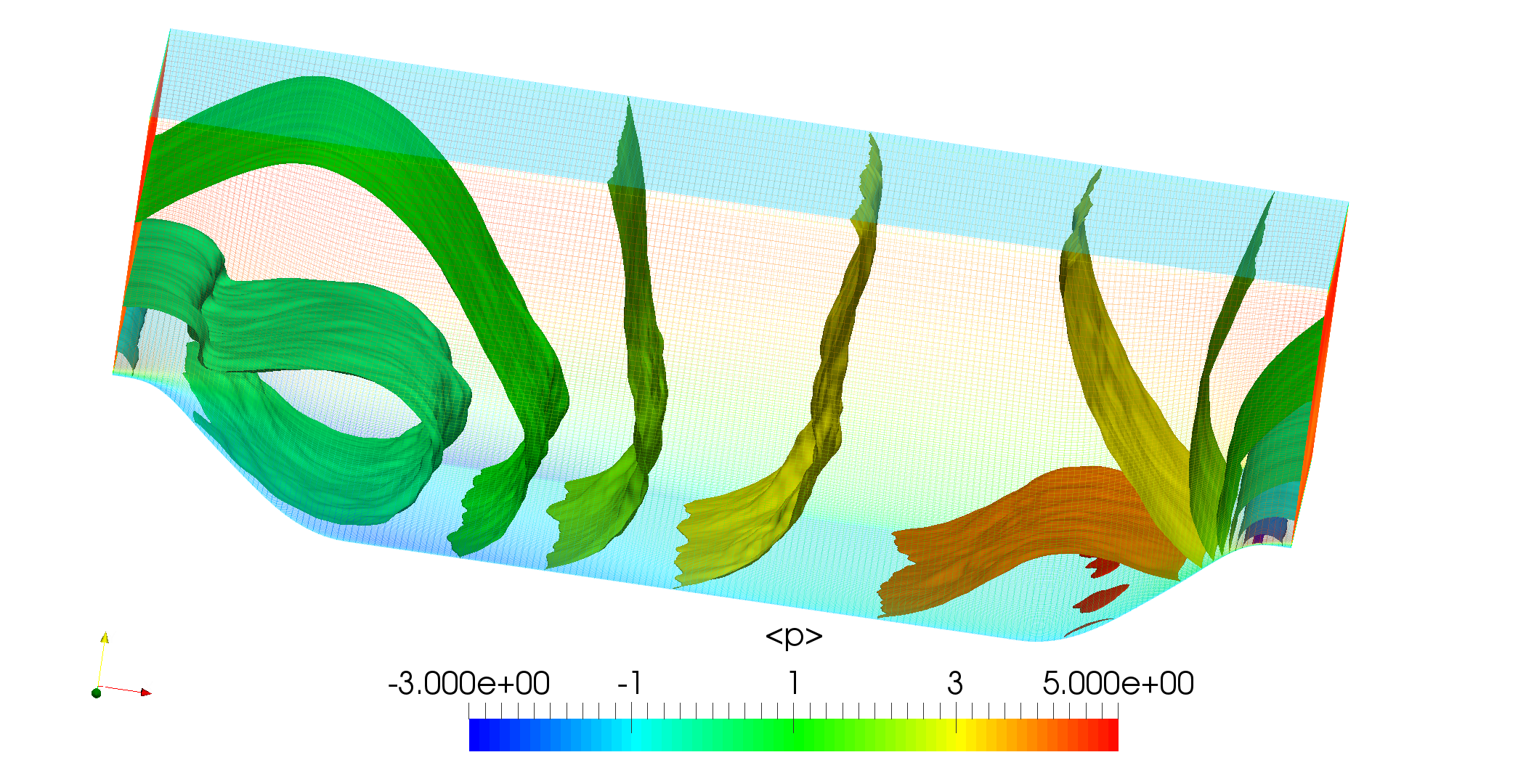}}
			\label{fig:pMeanLESG4}
		}
		\subfigure[LES (G4)]{
			\resizebox*{7cm}{!}{\includegraphics[width=\textwidth]{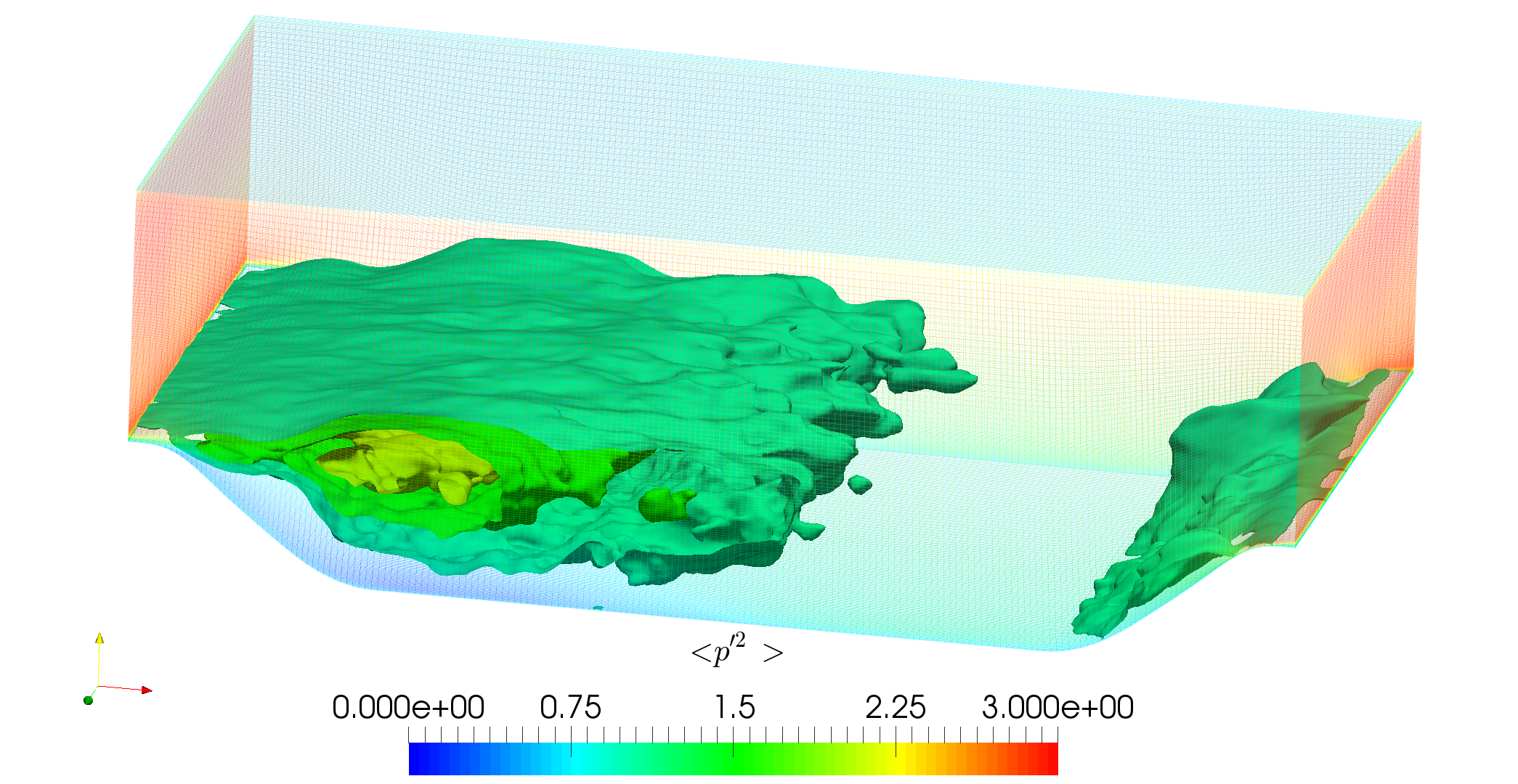}}
			\label{fig:pPrime2MeanLESG4}}

		\subfigure[NBSST (G3)]{
			\resizebox*{7cm}{!}{\includegraphics[width=\textwidth]{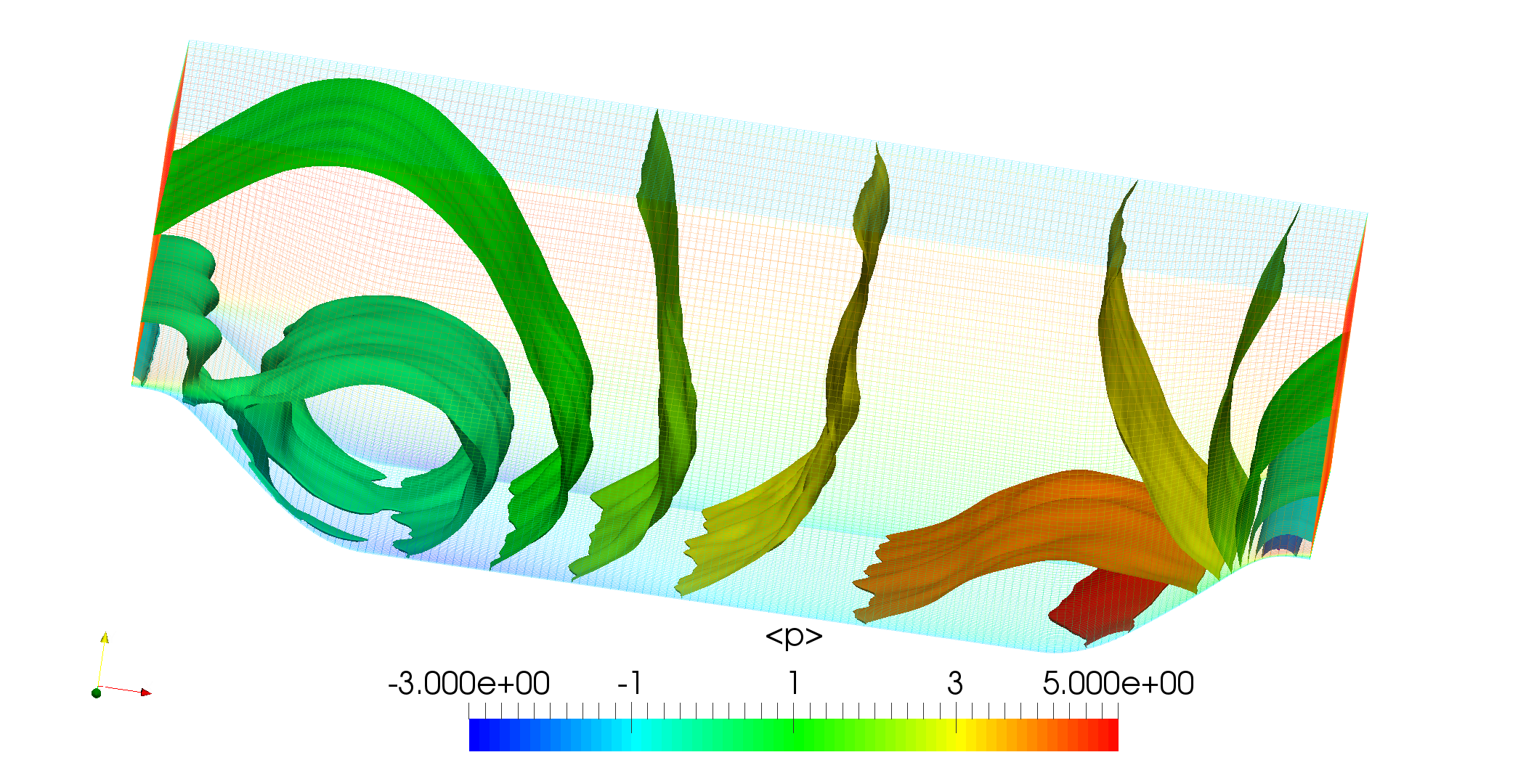}}
			\label{fig:pMeanNBSSTG2}
		}
		\subfigure[NBSST (G3)]{
			\resizebox*{7cm}{!}{\includegraphics[width=\textwidth]{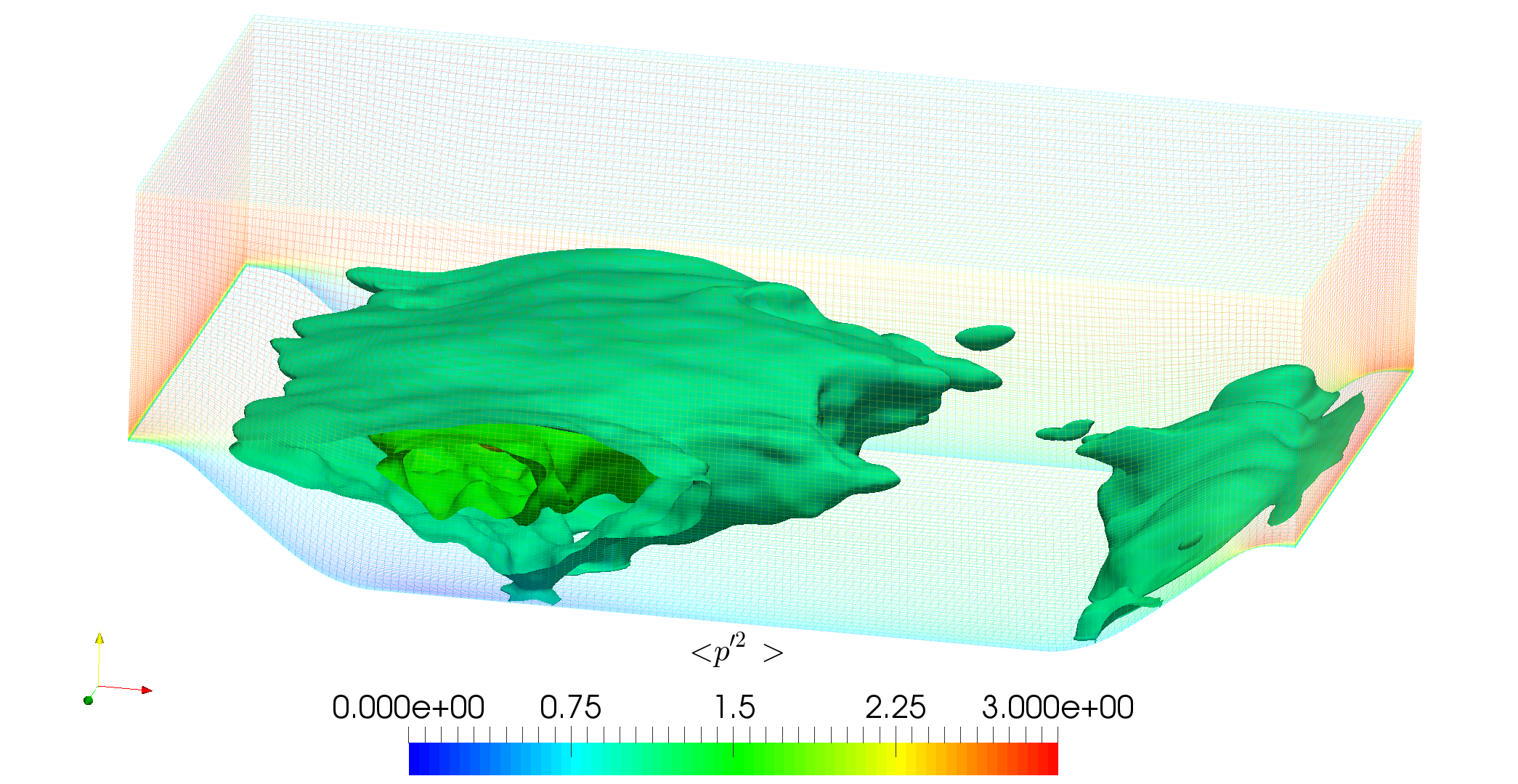}}
			\label{fig:pPrime2MeanNBSSTG2}}
	\end{minipage}
	\caption{Mean pressure and mean square of pressure fluctuation}
	\label{pressure}
\end{figure}

Figure \ref{pressure} shows the comparison of Mean and mean square of pressure fluctuations between LES simulation on G4 and NBSST simulation on G3 grid. Mean pressure comparison between the two cases does not show any significant difference, however comparison of mean square of pressure fluctuations shows considerable difference at the top of the hill near separation point and in the middle of the domain near reattachment point where fluctuating pressure is less than LES computation.

\subsection{Coherent Structures}
\begin{figure}
	\centering
	\begin{minipage}{150mm}
		\subfigure[LES (G4)]{
			\resizebox*{7cm}{!}{\includegraphics[width=\textwidth]{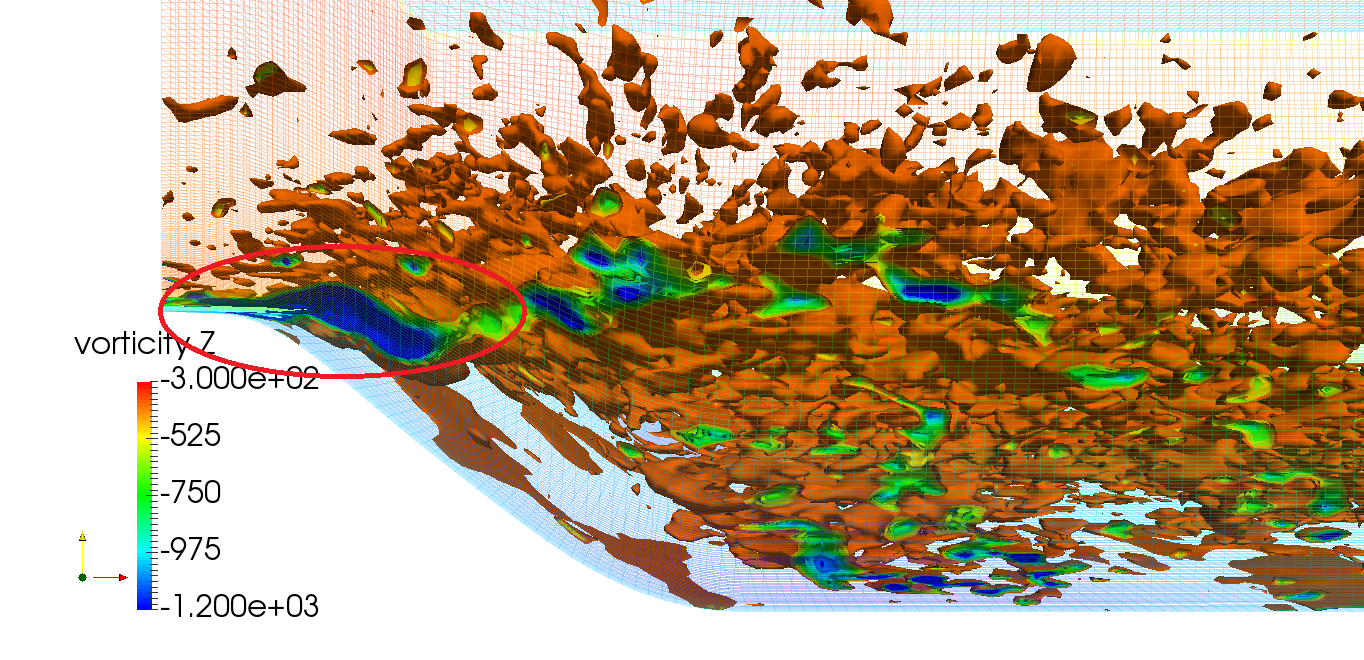}}
			\label{fig:Omegaz1}
		}
		\subfigure[LES (G4)]{
			\resizebox*{7cm}{!}{\includegraphics[width=\textwidth]{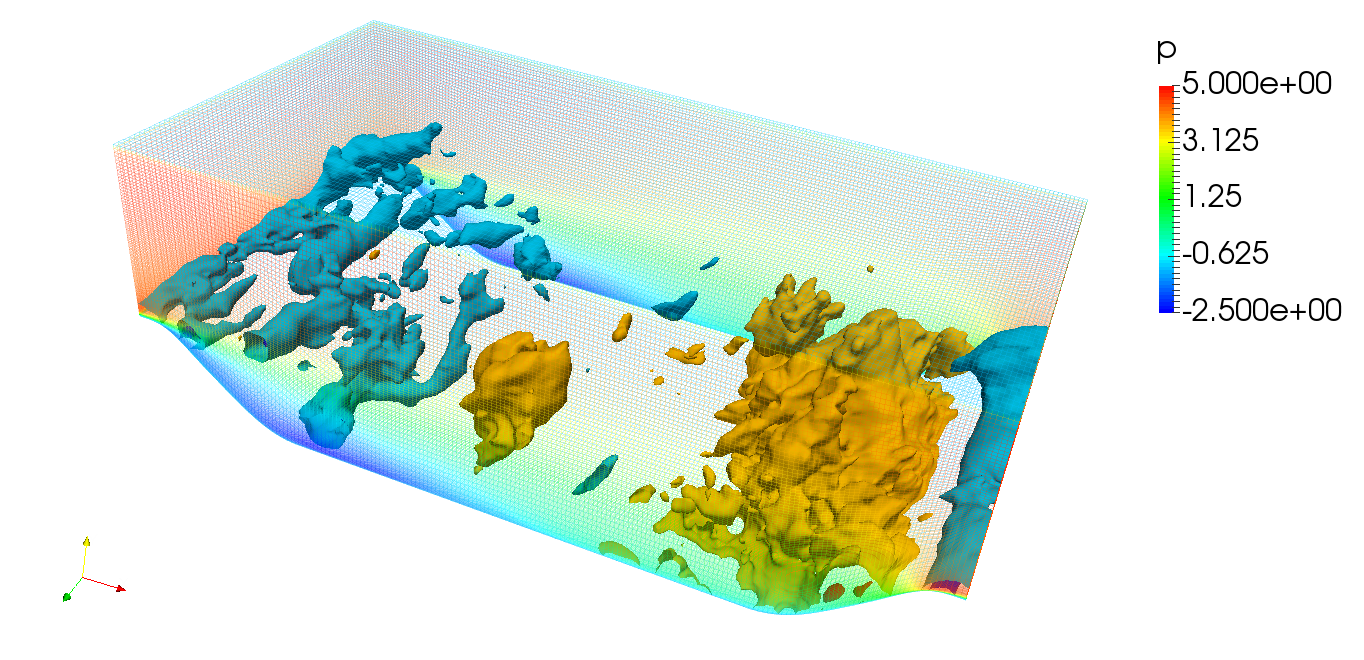}}
			\label{fig:p1}}
		
		\subfigure[LES (G3)]{
			\resizebox*{7cm}{!}{\includegraphics[width=\textwidth]{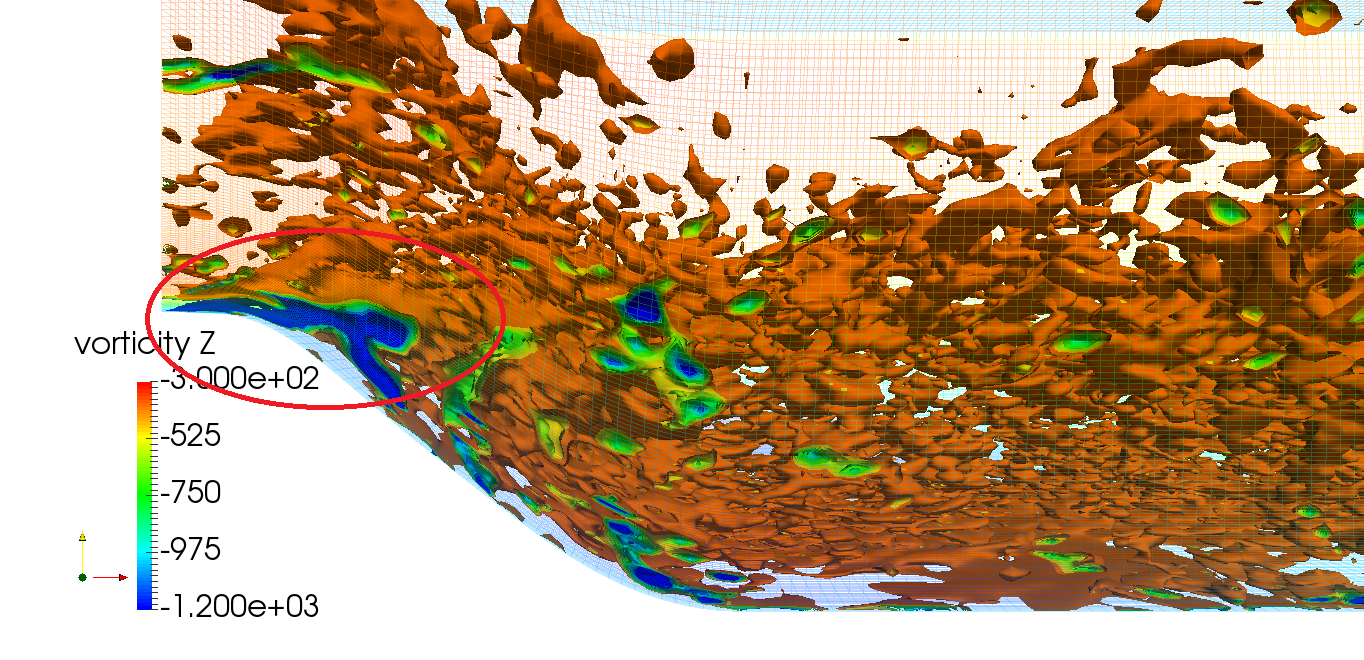}}
			\label{fig:Omegaz2}
		}
		\subfigure[LES (G3)]{
			\resizebox*{7cm}{!}{\includegraphics[width=\textwidth]{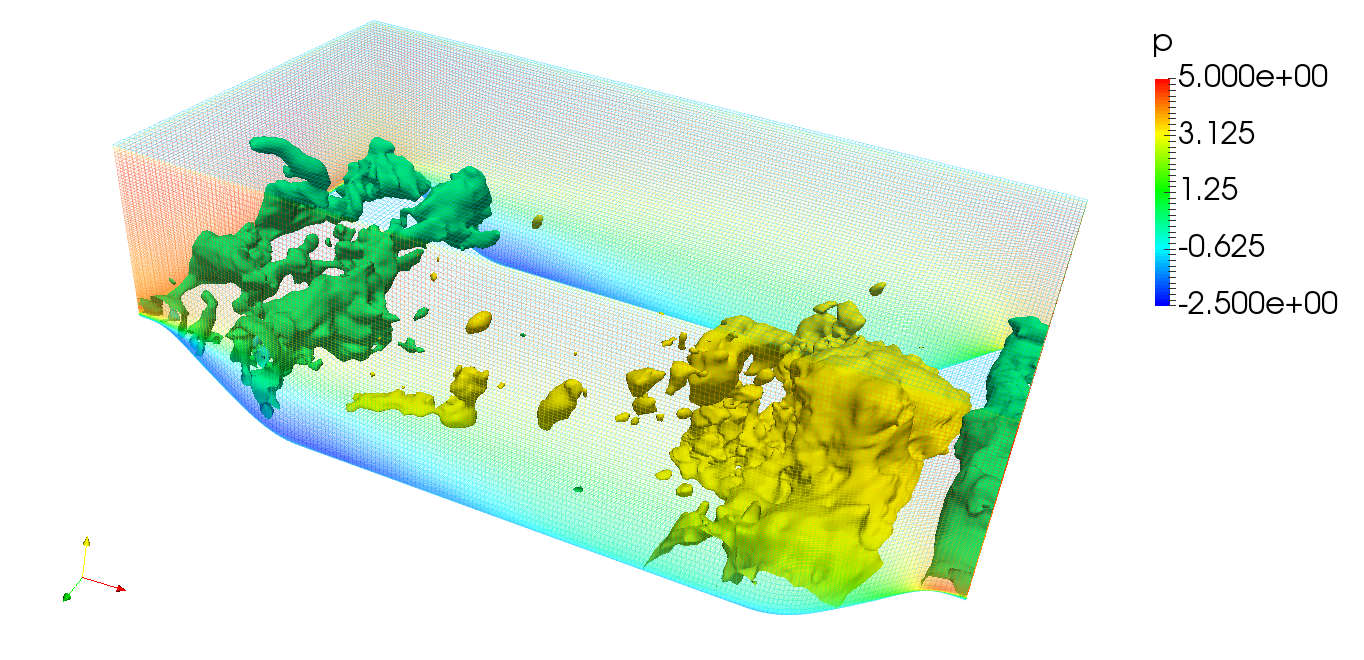}}
			\label{fig:p2}}

		\subfigure[NBSST (G3)]{
			\resizebox*{7cm}{!}{\includegraphics[width=\textwidth]{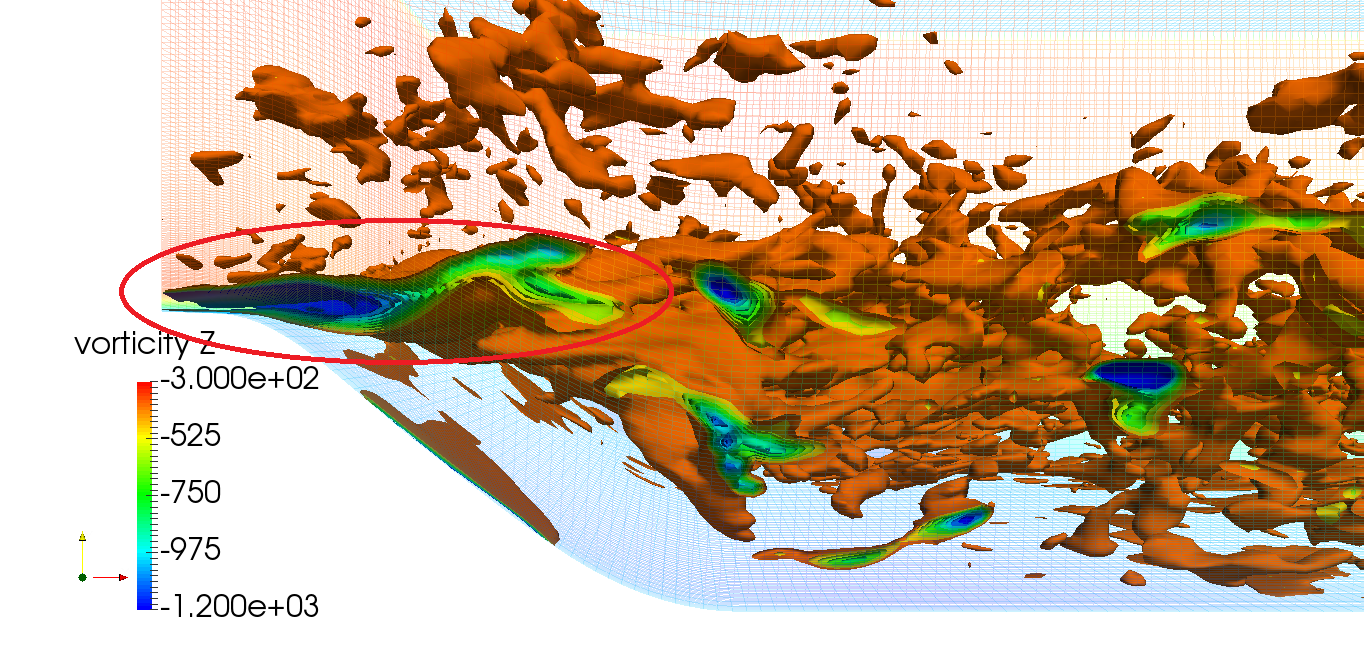}}
			\label{fig:Omegaz3}
		}
		\subfigure[NBSST (G3)]{
			\resizebox*{7cm}{!}{\includegraphics[width=\textwidth]{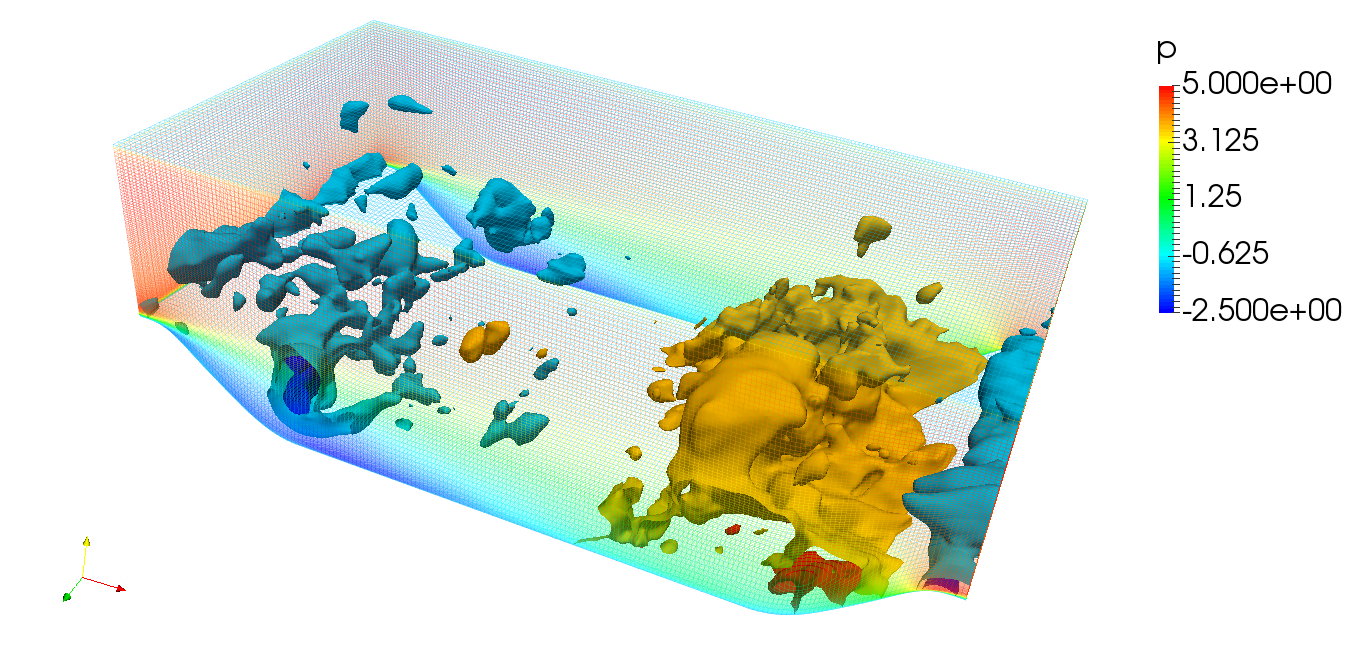}}
			\label{fig:p3}}
	\end{minipage}
	\caption{Instantaneous Flow Fields at Re=10595}
	\label{coherent}
\end{figure}

\begin{figure}
\centering
\includegraphics[width=\textwidth]{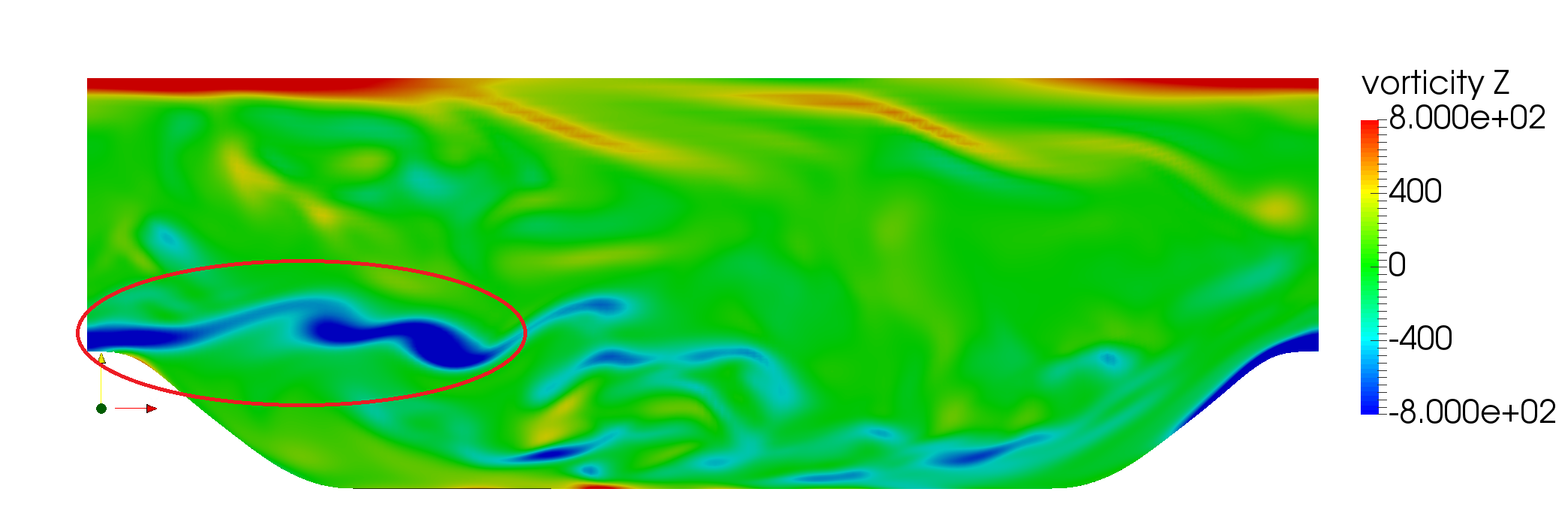}
\caption{Instantaneous spanwise vorticity using LES at Re=600}
\label{Re600}
\end{figure}

In Figures \ref{fig:Omegaz1}, \ref{fig:Omegaz2} and \ref{fig:Omegaz3}, iso-contours of spanwise vorticity component are shown.  In the previous studies \citep{frohlich2005highly}, existence of Kelvin-Helmholtz vortices and Gortler vortices at the top of the hill along with Helical pairing of vortices in the middle of the domain has been suggested at Re = 10595. Due to very small turbulent length and time scales present in these simulations due to high Reynolds number, only K-H vortices can be identified at the top of the hill at this Reynolds number as marked in Fig \ref{coherent}. However, K-H vortices can be very easily identified at low Re cases such as Re=600 as shown in Figure \ref{Re600}. In comparison to iso-contours of spanwise vorticity for LES simulation on G4, LES on G3 shows smaller length of the vortex instability region as soon as it breaks into turbulent eddies. However, NBSST simulation on G3 shows comparable length of vortex instability region and then breaks into turbulent eddies. 

Similar trend is observed in Figures \ref{fig:p1}, \ref{fig:p2} and \ref{fig:p3}, where pressure iso-contours are plotted for three simulations. The pressure rolls built up from the top of the hill disappears quickly in case of LES on G3 grid as compared to LES on G4 grid. However, it follows up to a similar distance for NBSST simulation on G3 grid. 

Another point to note is the size of smaller eddies present in LES simulations on G4 and G3 grids are similar whereas NBSST shows paucity of the smaller structures probably due to RANS modeling in the near wall region where length scales used for modelling of small turbulent structures is much larger than length scales used in LES. This is one of the reasons for better performance (poor performance) of hybrid methods for separated flows (attached flows). This suggests that care is required in the use of hybrid methods for attached and weakly separated flows because this issue will not be observed for massively separated flows. This also provides an advantage to the users to obtain similar coherent structures using hybrid RANS-LES model on a coarser grid.

\subsection{Proper Orthogonal Decomposition Analysis}
Instantaneous flow field provides insight into the flow structures, but it is also important for a model to predict the dynamics correctly. Here, POD has been used to asses the two-point statistical difference between predictions of these three simulations.  POD calculations have been performed over 201 snapshots taken at regular intervals in a period of approximately 8 flow through times (= 8*(Length of domain/average bulk velocity over the domain)). Flow fields have been decomposed into 201 eigen-values and eigen-functions corresponding to instantaneous pressure and instantaneous velocity field.  

\begin{figure}
	\centering
	\begin{minipage}{150mm}
		\subfigure[Pressure based POD (p)]{
			\resizebox*{7cm}{!}{\includegraphics[width=\textwidth]{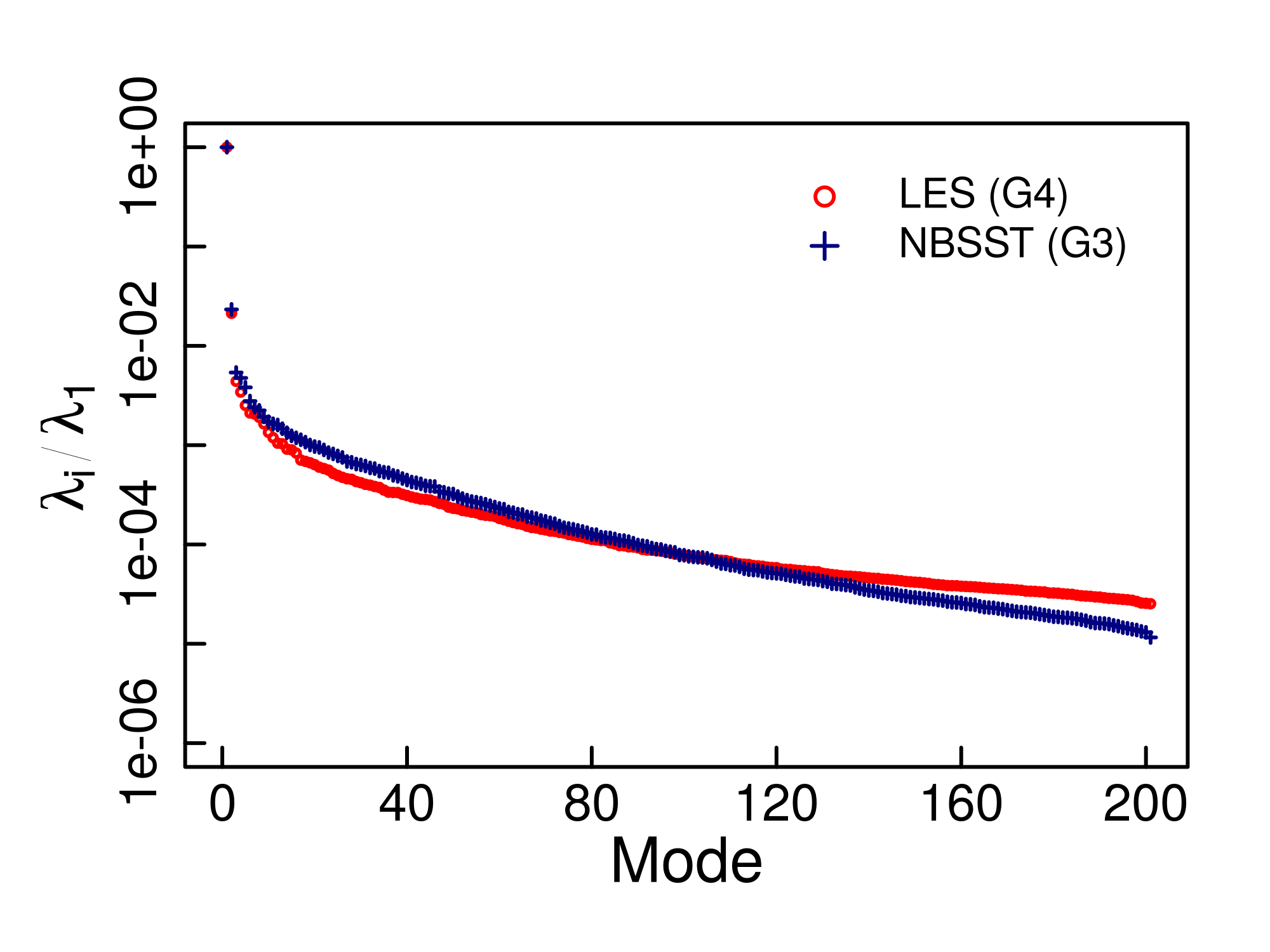}}
			\label{fig:peigen}
		}
		\subfigure[Velocity based POD ($\vec{V}$)]{
			\resizebox*{7cm}{!}{\includegraphics[width=\textwidth]{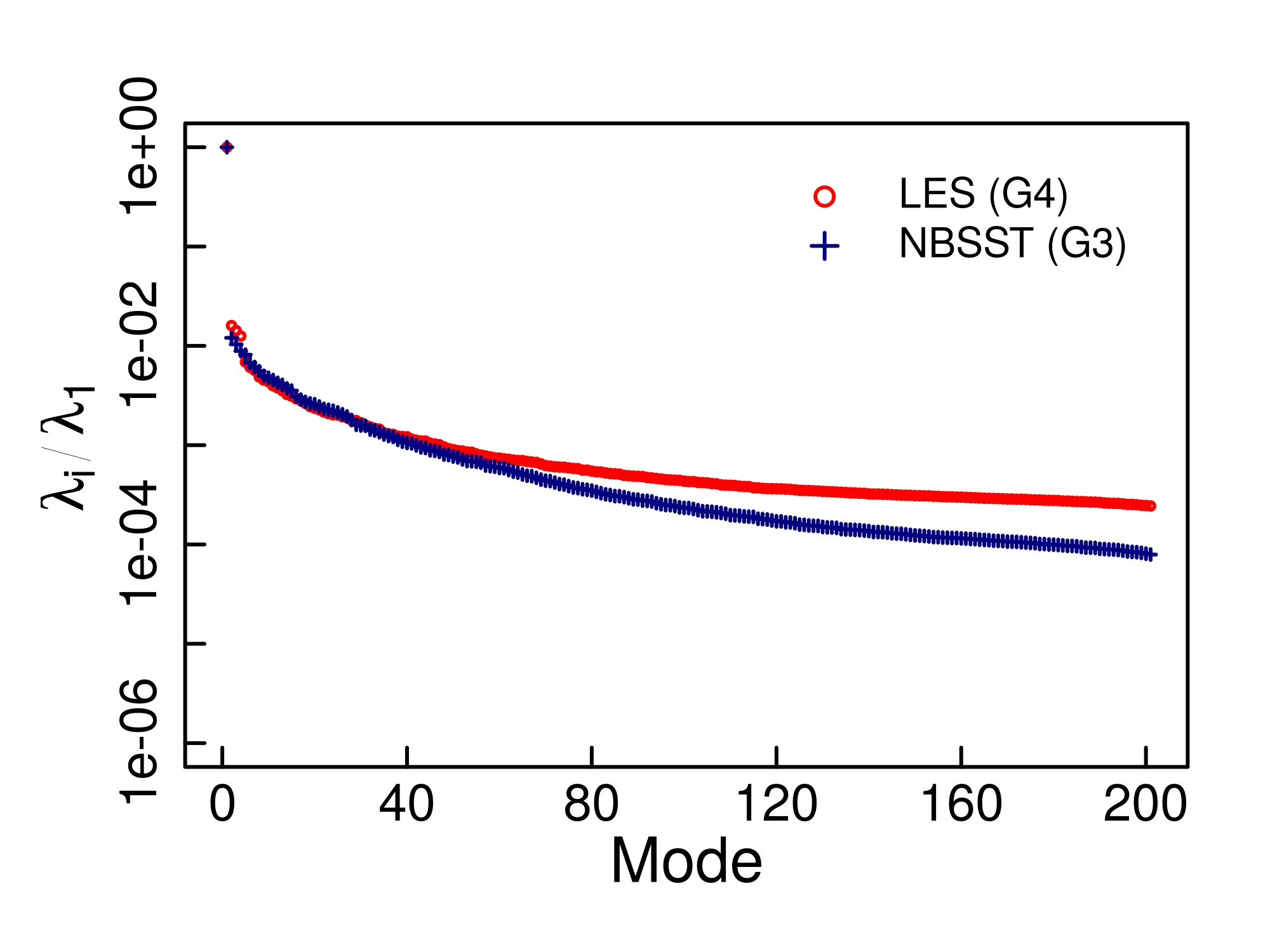}}
			\label{fig:Ueigen}}
%
			\end{minipage}
	\caption{Eigen values corresponding to Proper Orthogonal decomposition modes of pressure and velocity vector. Eigen-values are normalised with the eigen-value of first mode of the field itself. Here, first POD mode corresponds to the mean field.}
    \label{eigenValues}
\end{figure}

Let us consider a scalar field $\phi(x,t_n)$ with finite number of snapshots (here $n={1,2,...,201}$) 
Then, two-point auto-covariance matrix of dimension $n \times n$ can be constituted using ensamble $(< . >)$ of the scalar field over the spatial co-ordinates as:
\begin{equation}
\begin{split}
R(t_m,t_n) &= <\phi(x,t_m) . \phi(x,t_n)> \\
 &= \Sigma_k \lambda_k \phi_k(x)*\phi_k(x) \\
 \phi(x,t_n) &= \Sigma_k a_k \phi_k(x) \\
 where, <a_k a_{k'}> &= \delta_{kk'} \lambda_k
\end{split}
\end{equation}
Here, $\phi_k(x)$ are the empirical eigen-functions, $\lambda_k$ are the eigen-values corresponding to $k^{th}$ eigen-function and $a_k$ are the time coefficients corresponding to $k^{th}$ eigen-function. Since, two-point correlation matrix is constituted by the ensemble of square of instantaneous field over the domain, $k^{th} $eigen-values corresponds to the partition of energy associated with $k^{th}$ eigen-function. Therefore, the list of eigen-values has been arranged in decreasing order, so that first mode corresponds to the eigen-function containing maximum energy. Since, the correlation matrix is computed using instantaneous field, first eigen-function obtained is the time averaged mean of the field and corresponding eigen-value correspond to the energy associated with the mean flow. So, for easier representation, eigen-values have been normalized with the eigen-value of the first mode, so that, any eigen-value in figure  \ref{eigenValues} represents the energy associated with that eigen-function with respect to Mean flow energy. The same analysis can be performed for vector field such as velocity field, by considering a new scalar field formed by concatenating the three components in a sequence.

\begin{figure}
	\centering
	\begin{minipage}{150mm}
		\subfigure[Mode1(LES(G4))]{
			\resizebox*{7cm}{!}{\includegraphics[width=\textwidth]{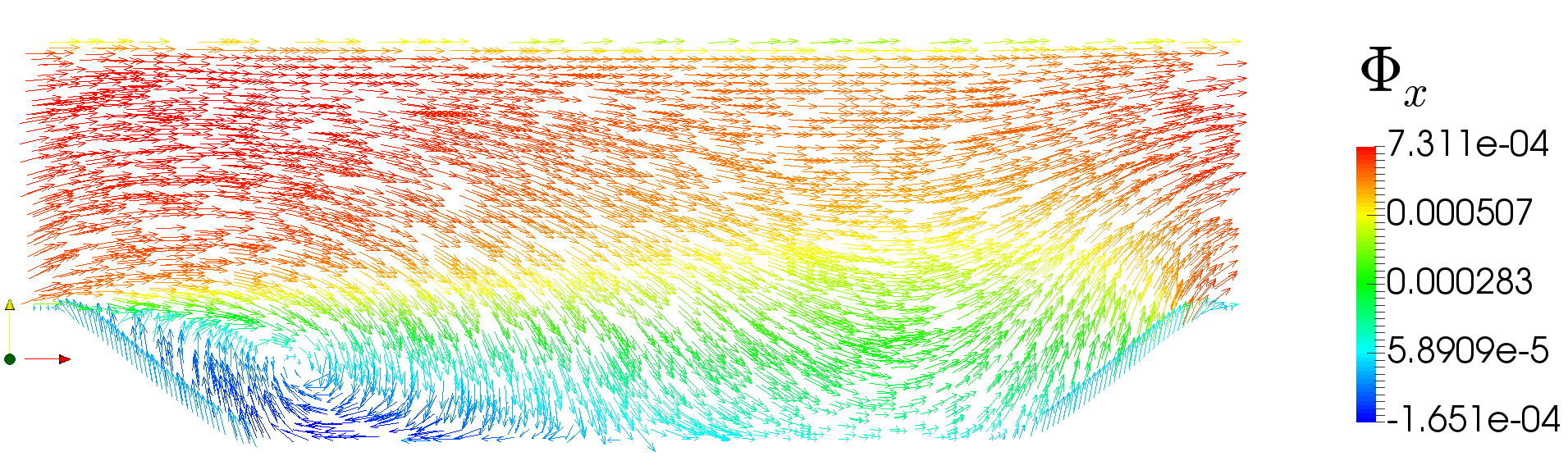}}
			\label{fig:Mode1LESG4U}
		}
		\subfigure[Mode1(NBSST(G3))]{
			\resizebox*{7cm}{!}{\includegraphics[width=\textwidth]{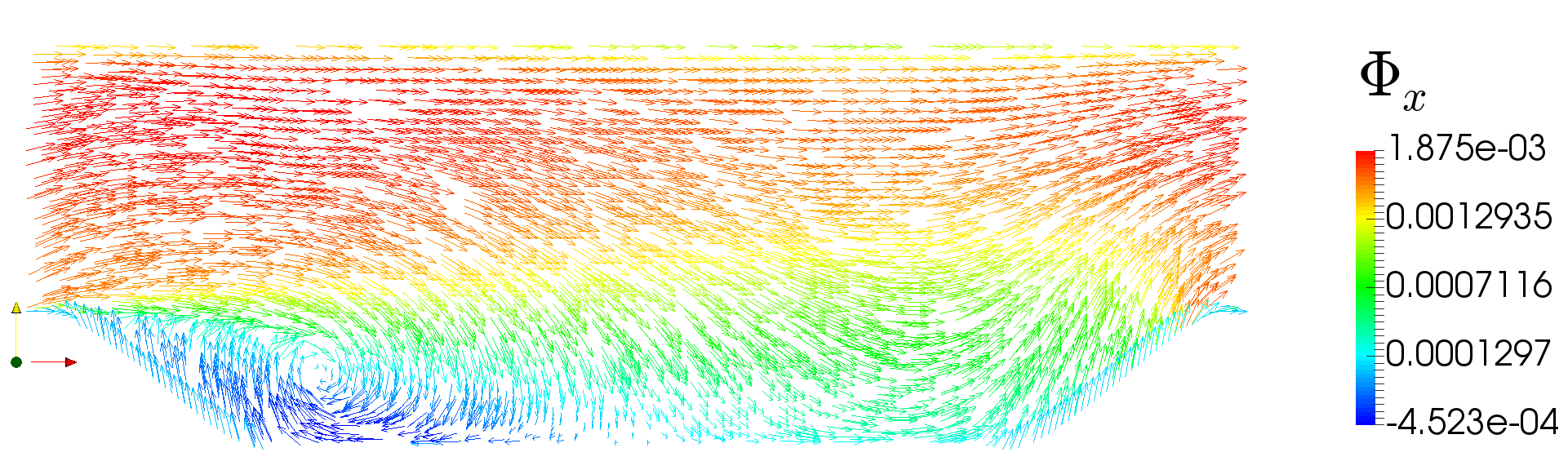}}
			\label{fig:Mode1NBSSTG2U}}
	\end{minipage}
	\caption{Two-dimensional vector plot for first eigen-mode for velocity based POD.}
            \label{Ueigenmode1}
\end{figure}

\begin{figure}
	\centering
	\begin{minipage}{150mm}		
		\subfigure[Mode1(LES(G4))]{
			\resizebox*{7cm}{!}{\includegraphics[width=\textwidth]{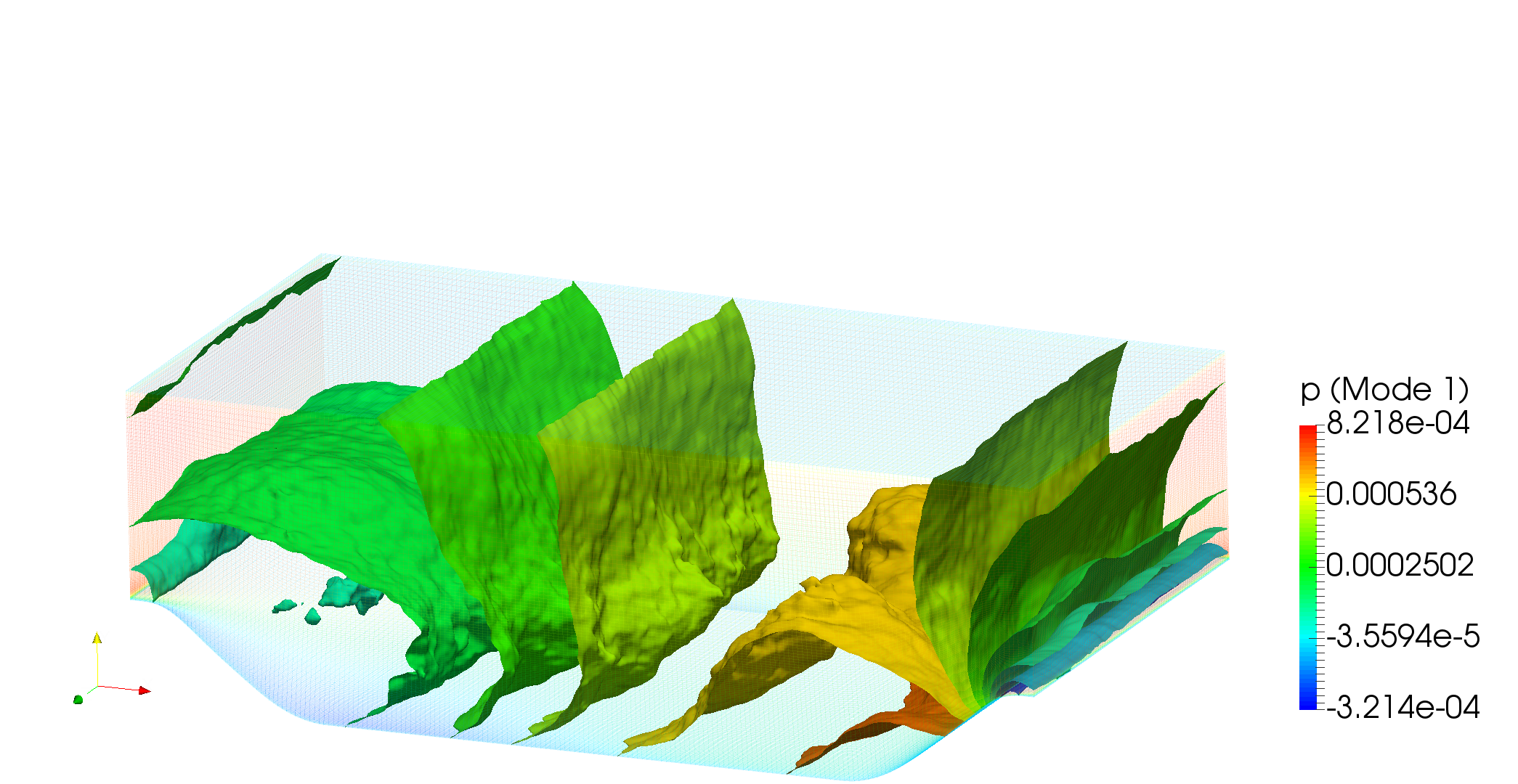}}
			\label{fig:Mode1LESG4p}
		}
		\subfigure[Mode1(NBSST(G3))]{
			\resizebox*{7cm}{!}{\includegraphics[width=\textwidth]{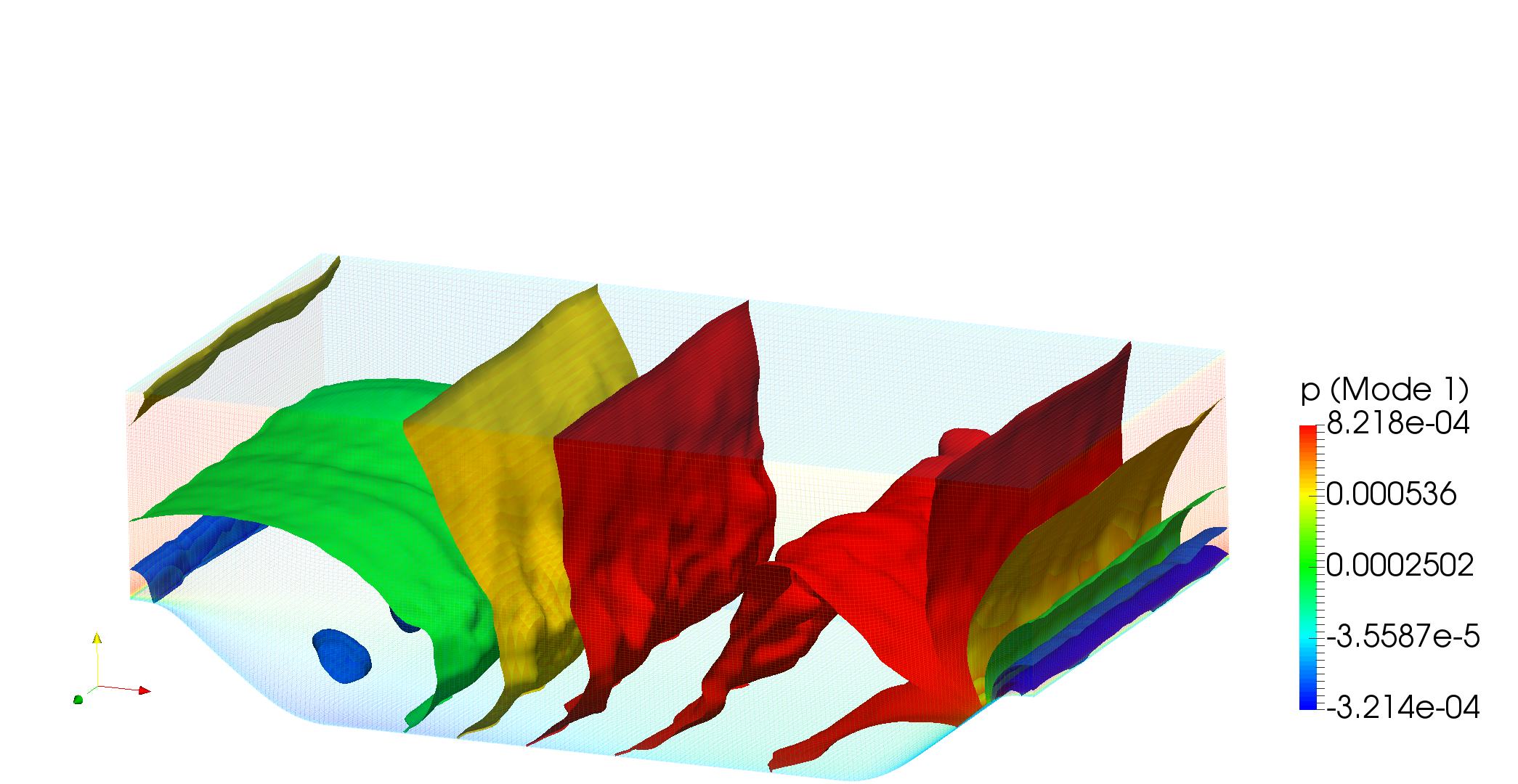}}
			\label{fig:Mode1NBSSTG2p}}
			
		\subfigure[Mode2(LES(G4))]{
			\resizebox*{7cm}{!}{\includegraphics[width=\textwidth]{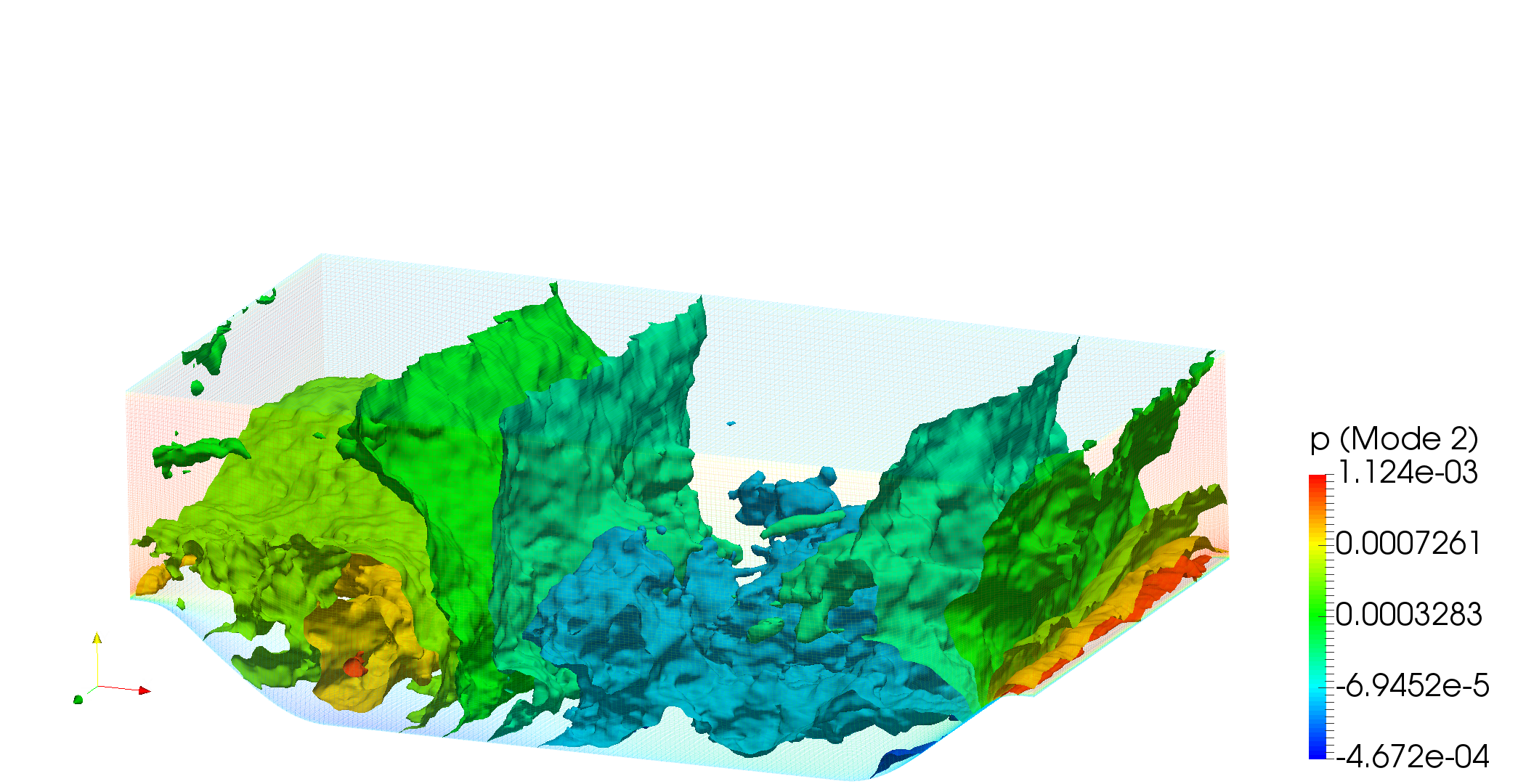}}
			\label{fig:Mode2LESG4p}
		}
		\subfigure[Mode2(NBSST(G3))]{
			\resizebox*{7cm}{!}{\includegraphics[width=\textwidth]{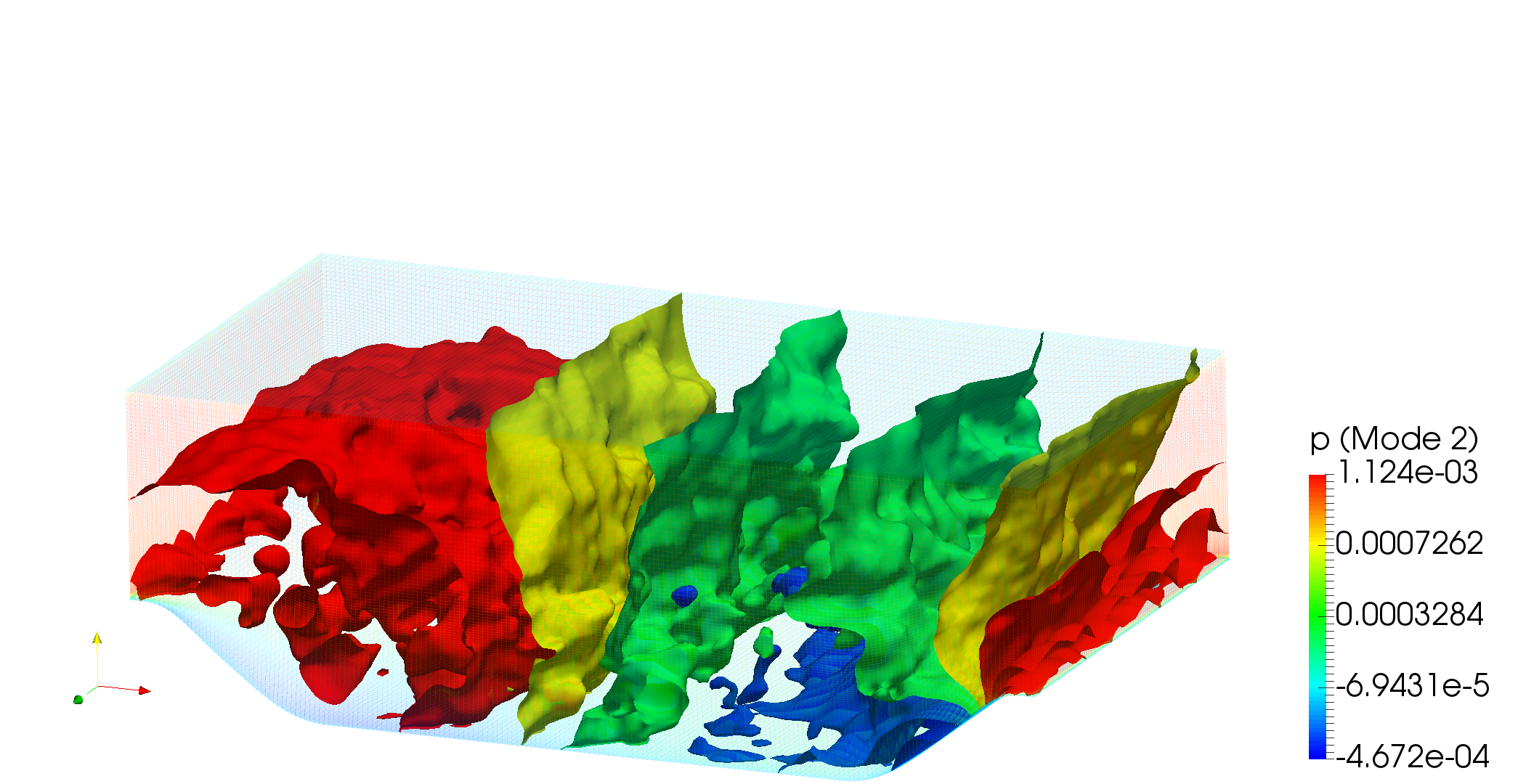}}
			\label{fig:Mode2NBSSTG2p}}	
			\end{minipage}
	\caption{Iso-contours for first and second eigen-mode for pressure based POD}
            \label{peigenmodes}
\end{figure}

LES simulation on G4 grid, which is very accurate due to high resolution of grid used for computation, has been used here as a reference for the assessment of the hybrid RANS-LES model simulation on G3 grid. In Figures \ref{fig:peigen} and \ref{fig:Ueigen}, eigen-values for instantaneous pressure and velocity fields have been plotted. Here, energy associated with mean mode of pressure POD is  92\% of the total energy whereas the energy associated with the mean mode of velocity field is 82\%. In Figure \ref{fig:peigen}, first two modes have much higher energy than the rest of the modes which follow continuous distribution of energy in higher modes; whereas in Figure \ref{fig:Ueigen}, only first mode has significantly higher energy whereas all the higher modes show gradual decrease in energy. In comparison with the eigen-values plot of LES simulation on G4 grid, the nonlinear BlendedSST on G3 grid has very close predictions of distribution of eigen-values for high energy modes and shows small difference in low energy modes. In figures \ref{fig:Mode1LESG4U} and \ref{fig:Mode1NBSSTG2U}, two dimensional vector plot for first eigen-mode of velocity based POD is shown. Both the plots show same flow structures, indicating NBSST simulation on G3 grid is capable of predicting the coherent structures accurately. Also, Iso-contours of first two modes of pressure based POD is shown in \ref{peigenmodes}. The structures seen in both the modes for NBSST is is found to be similar as LES simulation except magnitude difference. Since, these eigen-modes are vector and considered only to visualize the high energy coherent structures in the flow, the magnitude difference may not add too much value for the comparison.


\section{Conclusion}
In the current study, a comparison has been made to assess the solution inaccuracy due to lower grid resolution in Large Eddy Simulation (LES) and improvement of solution using Hybrid RANS-LES model. LES seems to provide very accurate results, for turbulent separating flows as in case of flow over periodic hills considered here, using highly resolved grid. However, the results deteriorates very rapidly for computations performed using same model on a coarser grid. At the same time, hybrid RANS-LES model (non-linear blended SST model considered here), provides significantly improved results than pure LES calculations on a coarse grid. The performance of the models have been assessed in terms of 1-point mean statistics, energy distributions and coherent structures in different modes of the flow obtained using proper orthogonal decomposition. 

LES and hybrid RANS-LES has a close agreement for the first and second order 1-point statistics while performing hybrid RANS-LES simulation on very coarse grid. Hence, the predictions of skin-friction and wall pressure coefficients and separation and reattachment locations show high dependence on grid in case of LES but the results converse for hybrid RANS-LES on a coarser grid. Since the flow quantities associated with the wall depends on complete flow information in the domain and can only be matched exactly if the flow structures and their dynamics are same in the domain. POD as an analysis tool for two-points statistics shows that hybrid RANS-LES model has close prediction of flow dynamics as compared to LES simulation on a highly refined grids. Also, looking at the computational cost, Hybrid RANS-LES simulation is very cheap as compared to the highly resolved LES simulation for the amount of inaccuracy we have to accept in Hybrid RANS-LES simulations. Hence, in all aspects, Hybrid RANS-LES model provides a good prediction of mean and instantaneous flow fields on a coarser grid compared to the LES simulation on highly refined grid.

\section*{Acknowledgments}
Simulations are carried out on the computers provided by the Indian Institute of Technology Kanpur (IITK) (www.iitk.ac.in/cc) and the manuscript preparation as well as data analysis has been carried out using the resources available at IITK. This support is gratefully acknowledged. This paper is an invited contribution for the $10^{th}$ International Symposium on Numerical Analysis of Fluid Flow and Heat Transfer - Numerical Fluids 2015, Rhodes, Greece.

\bibliographystyle{abbrvnat}
\bibliography{bibjot2014}

\section{Appendices}
\subsection{k-$\omega$ SST RANS Model}\label{appendix}
In this section details of the baseline RANS models are provided. K-$\omega$ SST model \citep{mentersst} is used as the baseline linear RANS model. The model solves two transport equation for the TKE and turbulent dissipation $\omega$  \citep{mentersst}
\begin{equation}
	\frac{\partial k}{\partial t}+\bar{u}_{j}\frac{\partial k}{\partial x_{j}}=P_{k}-\beta^{*}\omega k+\frac{\partial}{\partial x_{j}}\left[\left(\nu+\sigma_{k}\nu_{t}\right)\frac{\partial k}{\partial x_{j}}\right]
\end{equation}
\begin{align}
	\frac{\partial\omega}{\partial t}+\bar{u}_{j}\frac{\partial\omega}{\partial x_{j}}&=\frac{\gamma}{\nu_{t}}P_{k} 
	-\beta\omega^{2}+\frac{\partial}{\partial x_{j}}\left[\left(\nu+\sigma_{w}\nu_{t}\right)\frac{\partial\omega}{\partial x_{j}}\right]  
	\\ \nonumber &\qquad {}+ 2\left(1-F_{1}\right)\frac{\sigma_{w2}}{\omega}\frac{\partial k}{\partial x_{j}}\frac{\partial\omega}{\partial x_{j}}
\end{align}
where $P_{k}=\max(\tau_{ij}\partial\bar{u}_{i}/\partial x_{j},10\beta*\omega k$),
is the kinetic energy production term, $\beta^{*}=0.09$ is a model constant
and the last term in the $\omega$ equation is the cross-diffusion
term. $F_{1}$ is a blending function which has a value of one inside
the boundary layer and zero outside.  The limiting of the production term is an alternative to the use of damping function in the near-wall region. The turbulent stress tensor and viscosity are computed
in this model as follows
\begin{equation}
	\tau_{ij}=\frac{2}{3}k\delta_{ij}-2\nu_t \bar{S}_{ij}+N_{ij}
\end{equation}
\begin{equation}
	\nu_{t}=\frac{a_{1}k}{\max\left(a_{1}\omega,\sqrt{2\bar{S}_{ij}\bar{S}_{ij}}F_{2}\right)}
\end{equation}
Here $N_{ij}$ is the non-linear part of the turbulent stress tensor. $a_{1}=0.31$ is a model constant and $F_{2}$ is a blending function similar to $F_{1}$. The expressions for the blending functions are given by 
\begin{equation}
	F_{1}=\tanh\left[\left(\min\left\{ \max\left[\frac{\sqrt{k}}{\beta^{*}\omega d},\frac{500\nu}{d^{2}\omega}\right],\frac{4\sigma_{w2}k}{CD_{kw}d^{2}}\right\} \right)^{4}\right]\label{eq:kwsst_F1}
\end{equation}
\begin{equation}
	CD_{kw}=\max\left(2\frac{\sigma_{w2}}{\omega}\frac{\partial k}{\partial x_{i}}\frac{\partial\omega}{\partial x_{i}},10^{-10}\right)\label{eq:kwsst_cdkw}
\end{equation}
\begin{equation}
	F_{2}=\tanh\left[\left(2\frac{\sqrt{k}}{\beta^{*}\omega d},\frac{500\nu}{d^{2}\omega}\right)^{2}\right]\label{eq:kwsst_F2}
\end{equation}
The model constants are calculated by blending K-$\omega$ model near the wall and K-$\epsilon$ away from the
wall using the blending function $F_{1}$. The form is given by 
\begin{equation}
	\phi=\phi_{1}F_{1}+(1-F_{1})\phi_{2}\label{eq:kwsst_modelconstantblend}
\end{equation}
The model constants are given in Table \ref{Tab:sst_constants}.
For the non-linear K-$\omega$ SST model (NSST), a modified form of the 
non-linear constitutive relation proposed by Abe et al. \citep{abe_rans}
is used. The non-linear term is defined as follows
\begin{align}
	N_{ij}&= f_{NL}\frac{3\nu_{t}^{2}}{k}\left[f_{s}\left(2\bar{S}_{ik}\bar{S}_{kj}-\frac{2}{3}\bar{S}_{nk}\bar{S}_{kn}\delta_{ij}\right)-\bar{S}_{ik}\Omega_{kj}-\bar{S}_{jk}\Omega_{ki}\right] \nonumber \\
	&\qquad {} +2kd_{ij}^{w}
	\label{eq:Nij}
\end{align}
\begin{equation}
	f_{NL}=\frac{4}{3}C_{D}C_{B}(1-f_{w}(26))\label{eq:f_NL}
\end{equation}
\begin{equation}
	C_{B}=\frac{1}{1+22/3(C_{D}\nu_{t}/k)^{2}\Omega^{2}+2/3(C_{D}\nu_{t}/k)^{2}\left(\Omega^{2}-S^{2}\right)f_{B}}\label{eq:Cb}
\end{equation}
\[
f_{w}(\eta)=exp\left(-\left(y^{+}/\eta\right)^{2}\right)
\]
\begin{equation}
	f_{s}=1-\frac{S^{2}\left(\Omega^{2}-S^{2}\right)}{\left(\Omega^{2}+S^{2}\right)^{2}}\left\{ 1+C_{s2}C_{D}\left(\Omega-S\right)\frac{\nu_{t}}{k}\right\} \label{eq:fs}
\end{equation}
where $\Omega=\sqrt{\Omega_{ij}\Omega_{ij}}$ is the characteristic
rotation rate, $S=\sqrt{\bar{S}_{ij}\bar{S}_{ij}}$ is
the characteristic strain rate, $f_{B}=1+C_{\eta}C_{D}\nu_{t}/\left(k\left(\Omega-S\right)\right)$,
$y^{+}=u_{\tau}d_{y}/\nu$ is a non-dimensional wall distance, $\eta$
is a parameter, and $C_{D}=0.8$, $C_{\eta}=100$, and $C_{s2}=7$
are model constants. Finally, the expression for the term $d_{ij}^{w}$
can be written as 
\begin{eqnarray}
	d_{ij}^{w} & = & -\alpha_{w}f_{w}(26)\frac{1}{2}(d_{i}d_{j}-\frac{\delta_{ij}}{3}d_{k}d_{k}) \nonumber \\
	& + & f_{w}(26)(1-f_{r1}^{2})T_{d}^{2}\left\{ -\frac{\beta_{w}C_{w}}{1+C_{w}T_{d}^{2}\sqrt{S^{2}\Omega^{2}}}
	\left(\bar{S}_{ik}\Omega_{kj}-\Omega_{ik}\bar{S}_{kj}\right)\right\}  \nonumber \\
	& + & f_{w}(26)(1-f_{r1}^{2})T_{d}^{2}\left\{\frac{\gamma_{w}C_{w}}{1+C_{w}T_{d}^{2}S^{2}}\left(\bar{S}_{ik}\bar{S}_{kj}-\frac{\delta_{ij}}{3}S^{2}\right)\right\}
	\label{eq:dij_w}
\end{eqnarray}
where $d_{i}=\partial N_{i}/\partial x_{j}$, $N_{i}$ is the unit-normal,
and 
\begin{equation}
	f_{r1}=(\Omega^{2}-S^{2})/(\Omega^{2}+S^{2})\label{eq:nl_fr1}
\end{equation}
\begin{equation}
	T_{d}=\left\{ 1-f_{w}(15)\right\} k/\epsilon+f_{w}(15)\delta_{w}\sqrt{\nu/\epsilon}\label{eq:nl_Td}
\end{equation}
Here $\epsilon=\beta^{*}\omega k$ is the turbulence dissipation term.The model constants for the term $d_{ij}^{w}$ are given by 
\begin{equation}
	\alpha_{w}=1,\mbox{ }\beta_{w}=\frac{1}{4},\mbox{ }C_{w}=0.5,\mbox{ }\gamma_{w}=1.5,\mbox{ and }\delta_{w}=1.0\label{eq:dij_constants}
\end{equation}
The non-linear model NSST is used as the baseline model for the non-linear hybrid model. A damping function was used in Ref. \citep{abe_rans}. However, this term is not used in the current formulation as production limiter is used. In the original formulation of Abe et al. \citep{abe_rans}, the cross-diffusion term is not included. It has been included based on the recent improvement in results obtained with the inclusion of this term. 
\begin{table*}
	\centering                            \protect\caption{Model constants for the SST model.}                
	\begin{tabular}{|c|c|c|c|}
		\hline
		\multicolumn{4}{|c|}{$\phi_{1}$}\tabularnewline
		\hline
		$\sigma_{k}=0.85$ & $\gamma=5/9$ & $\beta=0.075$ & $\sigma_{w1}=0.5$\tabularnewline
		\hline
		\multicolumn{4}{|c|}{$\phi_{2}$}\tabularnewline
		\hline
		$\sigma_{k}=1.0$ & $\gamma=0.44$ & $\beta=0.0828$ & $\sigma_{w2}=0.856$\tabularnewline
		\hline
	\end{tabular}
	\label{Tab:sst_constants}
\end{table*}
\end{document}